\documentclass{article}
\usepackage[english]{babel}
\usepackage{amsfonts,amsmath,amssymb}
\usepackage{graphicx}

\usepackage[letterpaper,top=2cm,bottom=2cm,left=3cm,right=3cm,marginparwidth=1.75cm]{geometry}
\usepackage{enumerate}  
\usepackage{tikz}
\usepackage{bbold}
\usepackage{amsmath}
\usepackage{graphicx}
\usepackage{accents}
\usepackage[colorlinks=true, allcolors=blue]{hyperref}
\usepackage[linesnumbered,ruled,vlined]{algorithm2e}
\usepackage[skins]{tcolorbox}
\SetKwInput{KwInput}{Input}               
\SetKwInput{KwOutput}{Output} 
\usepackage{bm}
\usepackage{lipsum}
\usepackage{enumitem}
\setlist[enumerate]{parsep=4pt}
\usepackage[font=small]{caption}

\usepackage{amsthm}
\usepackage{natbib}
\setcitestyle{authoryear}

\usepackage[normalem]{ulem}

\newtheorem{example}{Example}
\newtheorem{theorem}{Theorem}

\newtheorem{lemma}{Lemma}
\newtheorem{assumption}{Assumption}

\newtheorem{remark}{Remark}
\usepackage{mathtools}  
\mathtoolsset{showonlyrefs}  
\usepackage{xcolor}

\usepackage{scalerel,stackengine}
\usepackage{tikz}
\usepackage{pgfplots}
\stackMath
\newcommand\reallywidehat[1]{%
\savestack{\tmpbox}{\stretchto{%
  \scaleto{%
    \scalerel*[\widthof{\ensuremath{#1}}]{\kern-.6pt\bigwedge\kern-.6pt}%
    {\rule[-\textheight/2]{1ex}{\textheight}}%
  }{\textheight}%
}{0.5ex}}%
\stackon[1pt]{#1}{\tmpbox}%
}
\title{Identifying Sparse Treatment Effects in High-dimensional Outcome Spaces}
\author{Yujin Jeong, Emily Fox, Ramesh Johari}
\date{}

\begin{document}

\maketitle

\begin{abstract}    Based on technological advances in sensing modalities, randomized trials with primary outcomes represented as high-dimensional vectors have become increasingly prevalent. For example, these outcomes could be week-long time-series data from wearable devices or high-dimensional neuroimaging data, such as from functional magnetic resonance imaging. This paper focuses on randomized treatment studies with such high-dimensional outcomes characterized by sparse treatment effects, where interventions may influence a small number of dimensions, e.g., small temporal windows or specific brain regions. Conventional practices, such as using fixed, low-dimensional summaries of the outcomes, result in significantly reduced power for detecting treatment effects. To address this limitation, we propose a procedure that involves subset selection followed by inference. Specifically, given a potentially large set of outcome summaries, we identify the subset that captures treatment effects, which requires only one call to the Lasso, and subsequently conduct inference on the selected subset. Via theoretical analysis as well as simulations, we demonstrate that our method asymptotically selects the correct subset and increases statistical power. 
\end{abstract}

\section{Introduction}\label{sec:introduction}

Due to technological advances in sensing modalities, randomized controlled trials with primary outcomes represented as high-dimensional vectors have become increasingly prevalent; examples include time series data from wearable sensors (e.g., heart rate or glucose levels), or neuroimaging sensor data.  When trial outcomes are high-dimensional, the scientist faces a challenge: in which dimension(s) is there a treatment effect, if any?  On one hand, the scientist could pre-commit to a small number of dimensions and test for treatment effects there (e.g., a fixed window of time in the wearables setting).  However, this risks missing the treatment effect.  On the other hand, if the entire set of dimensions is considered, then finding treatment effects amounts to searching for a needle in a haystack.

In this paper, we consider the problem of causal inference from experimental data with high-dimensional outcomes.  Our emphasis is on the development of a practical technique that allows the scientist to consider high-dimensional outcome summaries, and yet successfully identify and estimate sparse treatment effects.  We present a practical, easily deployed method that identifies the subset of outcome summaries that captures the sparse treatment effect given a large pool of outcome summaries, and then carries out inference for the treatment effect on the selected subset. 

In the remainder of this section, we present some motivation for our approach with examples from real-world and semi-synthetic experiments. 
First, we observe that for interpretability, instead of working with the raw high-dimensional outcomes, scientists often study a compressed summary of the raw data. 
A common practice is to compress the outcomes into a one-dimensional summary of interest, $Y \in \mathbb{R}$; alternatively, one can consider $p$ one-dimensional summaries, concatenated as $Y \in \mathbb{R}^p$, where $p$ is small, and employ multiple testing. This practice is illustrated in the examples below. 

\begin{enumerate}[resume,listparindent=1.5em,  labelsep=2em]
    \item \textbf{Time series data from wearable devices.} 
    Over the last decade, continuous glucose monitoring (CGM) has become an essential component of diabetes management for many people with type 1 diabetes. The study by \cite{aleppo2017replace} investigates whether the use of CGM without confirmatory blood glucose monitoring (BGM) measurements is as safe and effective as using CGM adjunctive to BGM in adults with well-controlled type 1 diabetes through a randomized controlled experiment. Their primary outcome was time-in-range (the percentage of time when the patient's blood glucose level lies between 70–180 mg/dL) over the 26-week period of monitoring. In this study, the raw outcome is a time series of glucose measurement over 26 weeks, and their choice of one-dimensional outcome summary $Y$ is time-in-range over the 26-week period.

    \item \textbf{Neuroimaging data.} Neuroscientists often investigate the impact of external stimuli on brain activity using neuroimaging techniques, such as functional magnetic resonance imaging (fMRI). In such cases, outcomes are high-dimensional in the form of images or a time series of images. 

    For example, \cite{koenig2017effects} studied the effects of the insulin sensitizer metformin on Alzheimer’s disease through a randomized placebo-controlled crossover design.
    Arterial Spin Label MRI data, which measures cerebral blood flow (CBF) in different brain regions, was collected at week 0 as baseline and at week 8. 
    Then the researchers tested whether the treatment had significant effects on CBF in 21 pre-defined regions of interest. In this study, the raw outcome is a pre-processed brain image and their choice of outcome summaries, $Y \in \mathbb{R}^{21}$, is measured CBF in 21 regions of interest. 

    In another example, \cite{anand2005antidepressant} studied the effects of the antidepressant sertraline on corticolimbic connectivity using a matched pairs experimental design. They obtained a series of fMRI scans (512 time points) at week 0 as baseline and week 6. Then the researchers compared correlation coefficients between the time series of four different regions of interest and tested whether there are significant differences between the control and treatment groups. In this study, the raw outcome is a time series of brain images and their choice of outcome summaries, $Y \in \mathbb{R}^{6}$, is a vector of correlation coefficients between four different regions of interest. 

\end{enumerate} 

This practice is common when scientists are interested in a few specific outcome summaries based on their scientific motivation. However, if scientists are interested in unraveling how the treatment effect may manifest in the high-dimensional outcome space, this practice may fail to capture the treatment effect when the treatment effect is sparse. 

\begin{figure}[t]
    \centering
    \includegraphics[scale = 0.55]{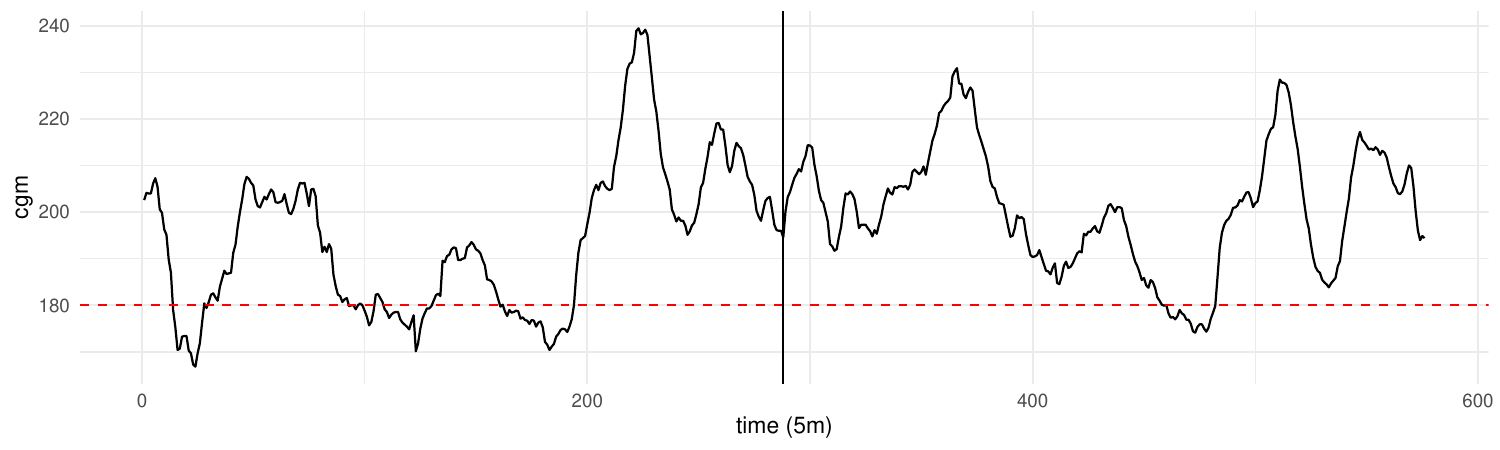}
    \caption{An example trace representing the daily average of CGM data in week 1 (before solid vertical line) and week 2 (after the solid vertical line).  The time series is recorded every 5 minutes, resulting in 24*12 = 288 points before and after the potential intervention time (solid vertical line).}
    \label{fig:cgm-example}
\end{figure}

To demonstrate this issue, we generate a semi-synthetic scenario to simulate realistic treatment effects observable in a clinical system for managing type 1 diabetes. We use CGM data from the ongoing 4T study at Stanford's Lucille Packard Children's Hospital \citep{prahalad2024equitable}.
In this study, patients are reviewed by clinicians weekly and, if an adjustment is deemed necessary, the clinician will send a message. These messages represent the interventions and often their effects may be time-localized, such as a correction to overnight hyperglycemia. To mimic this scenario, we generate synthetic treatment effects as follows. We randomly select 10,000 two-week long CGM traces, each providing recordings every 5 minutes, and retain the 5,000 traces with minimal missing data; see Section \ref{sec:semisynth} for details.
We treat the first week's data as a pre-treatment time series (i.e., pre-treatment covariates) and the second week's data as post-treatment. Each week's data is then averaged across days in the week in a time-aligned fashion to create a 24-hour average CGM representation that can be used to highlight the potential impacts of interventions on specific times of the day. An example CGM trace is shown in Figure \ref{fig:cgm-example}. We created treatment and control groups by randomizing treatment assignment with a 50$\%$ probability of receiving a treatment. To create the synthetic treatment effect, we localize an effect around lunchtime by reducing glucose levels of the treated units by 5 units between 12pm and 2pm, simulating a lunchtime intervention.

\begin{figure}[t]
    \centering
    \includegraphics[scale = 0.6]{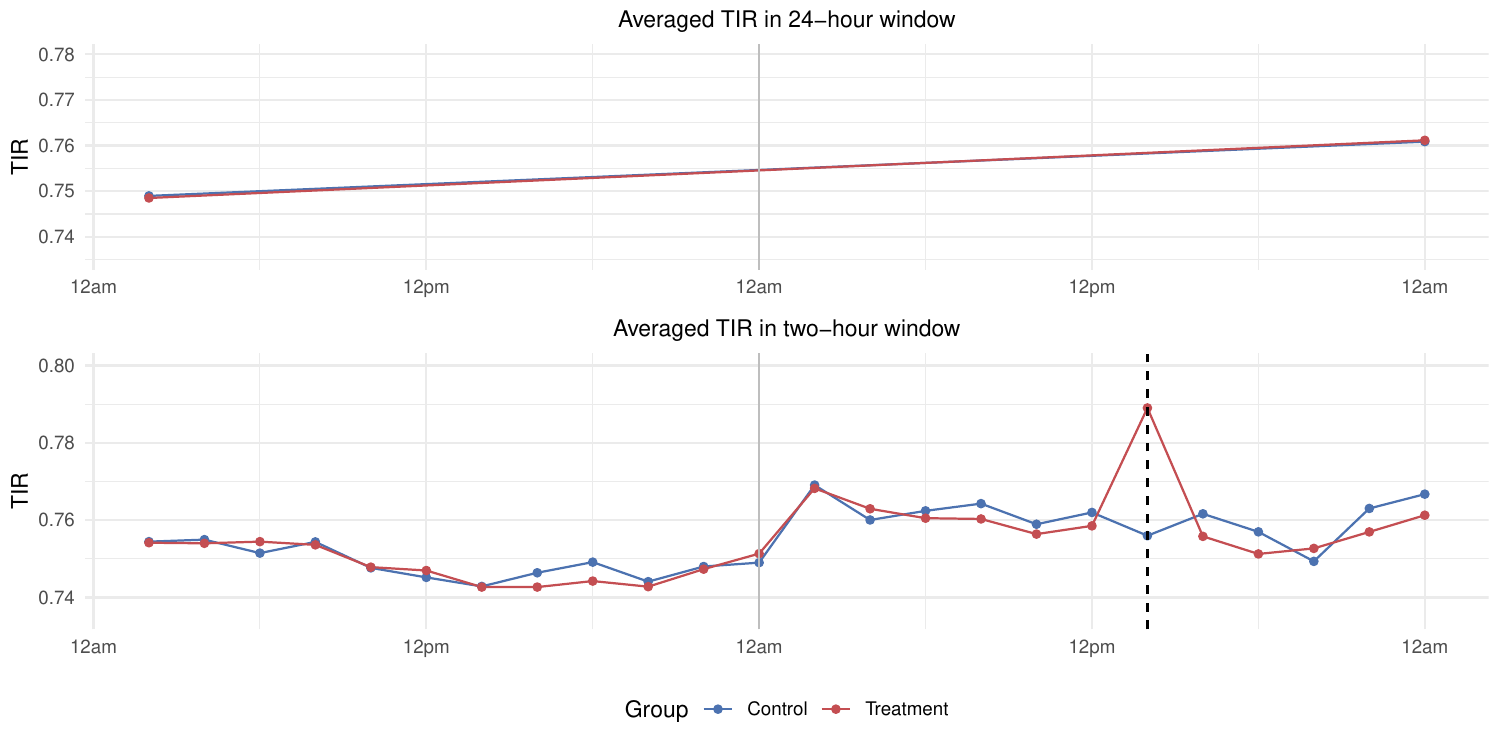}
    \caption{The averaged TIR for both the control group and the treatment group where there is a synthetic treatment effect applied between 12pm and 2pm. The dashed line indicates a lunchtime. The plot above shows that TIRs computed with 24-hour time windows fail to capture the treatment effect. On the other hand, the plot below implies that TIRs computed with two-hour time windows may capture the treatment effect.}
    \label{fig:TIR-motivation}
\end{figure}

If a scientist considers a one-dimensional summary, TIR (the percentage of time when the patient's blood glucose level lies between 70–180 mg/dL) over 24 hours, the treatment signal is lost as the treatment effect is sparse. This is illustrated in Figure \ref{fig:TIR-motivation}. On the other hand, TIR calculated with a two-hour time window seems to be able to capture the sparse treatment effect. In practice, however, the scientist might not know the duration and location of the actual treatment effect. 

The preceding discussion motivates the main challenges we address in this paper.  First, given a large number of possible outcome summaries (e.g., windows of different start time and duration), where is a treatment effect most likely present?  Second, what is an efficient means of leveraging pre-treatment covariates (e.g., the prior week's CGM time series) to improve detection of appropriate outcome summaries?  And third, what is the performance of such an approach, both in terms of recovery of the appropriate outcome summaries, as well as in terms of statistical power?

\subsection{Our Objective and Approach}

We consider a setting where the outcome of interest lies in the high-dimensional space $\mathbb{R}^p$, with $p$ possibly greater than $n$. As in the examples in the previous section, the outcome of interest may be $p$ one-dimensional summaries of the raw outcome. We assume that the treatment effect is sparse, meaning that the treatment effect manifests in only a few directions in the high-dimensional space $\mathbb{R}^p$. Additionally, the scientist does not know which dimensions effectively capture the treatment effect.

Our main contribution in this paper is a practical method for identification of the subset of outcome summaries in which a treatment effect manifests when the dimension $p$ is large. 
We propose a two-step procedure given two data splits. For example, in a clinical setting, we often begin with a small-scale experiment as pilot or proof-of-concept (POC) study, followed by a much larger clinical trial. Alternatively, these two data splits can also be obtained through sample-splitting. In the first stage, the data scientist defines a large pool of outcome summaries and selects a subset where the treatment effect may manifest using the first data split. In the second stage, the second data split is used to estimate the treatment effect on this selected subset.

Our method primarily focuses on the subset selection process: how to choose a subset of outcome summaries in which the treatment effect may manifest. We introduce a sparse regression approach for subset selection, where we employ a regularized weighted linear regression with carefully chosen weights. We show that our method can recover the subset of outcome summaries consistently. From simulated experiments, we demonstrate the effectiveness of our method, particularly when the treatment effect is relatively weak and where $p\gg n$ during the subset selection phase.  In our clinical applications of interest, we typically face $p \gg n$ during the initial POC study due to the small number of subjects.  In other settings, such as analysis of gene expression data, the massively high-dimensional outcomes create a $p \gg n$ scenario even when $n$ is still large enough to perform sample-splitting.

\subsection{Outline of the Paper}
In Section \ref{sec:related}, we discuss related work.  In Section \ref{sec:preliminaries}, we introduce the formal setting of the paper. In Section \ref{sec:baseline}, we provide an overview of a baseline approach for how a data scientist might select a subset of outcome summaries where the treatment effect potentially manifests. Section \ref{sec:proposed} outlines our proposed method based on sparse regressions. Section \ref{sec:wls} provides technical insights behind the proposed method and Section \ref{sec:recovery} presents theoretical results. We conclude in Section \ref{sec:num-exp} where we evaluate the performance of our proposed procedure on synthetic and semi-synthetic data sets.

\section{Related Work}
\label{sec:related}

This work considers the general randomized controlled setting in causal inference \citep{imbens2015causal}. In randomized controlled experiments, linear regression has been employed to improve the precision of average treatment effect estimators through linear regression adjustments with pre-treatment covariates \citep{deng2013improving, lin2013agnostic, ying2023variancereduction}. 
\cite{bloniarz2016Lasso} and \cite{wager2018highdimensional} have a discussion of high-dimensional regression adjustments when there are a large number of pre-treatment covariates. Our paper also studies high-dimensional regression techniques within randomized controlled experiments, but we focus on the setting where outcomes are high-dimensional instead of pre-treatment covariates.

One of the most popular high-dimensional linear regression techniques is the Lasso introduced by \cite{Tibshirani1996Lasso}. Conditions for oracle inequalities for the Lasso have been studied by  \cite{bunea2007c}, \cite{van2008high}, \cite{zhang2008sparsity}, \cite{meinshausen2009Lasso}, and \cite{bickel2009Lasso}. Furthermore, variable selection properties of the Lasso have been investigated by \cite{meinshausen2006high}, \cite{zhao2006model}, \cite{lounici2008sup-norm}, and \cite{wainwright2009sharp}. In addition, sample-splitting and subsequent statistical inference procedures have been developed in \cite{wasserman2009high} and \cite{meinshausen2009pvalues}.

The average treatment effect estimator investigates the mean difference between the treated group and the control group. Therefore, our work is also closely related to high-dimensional mean testing (the problem of testing whether a population mean $\mu$ equals to some known vector $\mu_0$ under the high-dimensional regime) \citep{huang2022overview}. Among the array of high-dimensional mean testing methods, our work is closely related to the projection test \citep{lopes2011powerful, huang2015projection, liu2022msptesting}. This approach involves mapping the high-dimensional samples to a lower-dimensional space and subsequently applying traditional methods, such as Hotelling's $T^2$ to the projected samples. Our paper adopts a causal inference perspective, exploring the interplay between treatment assignments, pre-treatment covariates, and post-treatment outcomes. Moreover, we employ a linear regression specification, which is more practical for practitioners to implement using tools such as \texttt{glmnet} in R.

\section{Preliminaries} \label{sec:preliminaries}

We consider a setting with $n$ individuals.  Each individual receives a single binary treatment of interest $T_i \in \{0, 1\}$, and we observe a subsequent vector of outcome summaries, denoted by $Y_i \in \mathbb{R}^p$, for each individual. In our setting, the primary outcome, a vector of outcome summaries, is high-dimensional: we allow the dimension $p$ to increase with $n$. In our subsequent presentation, we will also consider the inclusion of pre-treatment covariates $X_i \in \mathbb{R}^m$ for each individual.  

Following the potential outcomes framework \citep{imbens2015causal}, each individual has a $p$-dimensional pair of potential outcomes: $(Y_i(0), Y_i(1))$.  We take a super-population perspective where
$$
(X_i, Y_i(0), Y_i(1)) \stackrel{\text{i.i.d}}{\sim} \mathbb{P}.
$$  Moreover, we consider randomized controlled trials where $P(T_i = 1) = \pi$ (independently across $i$, with $0 < \pi < 1$) and $$T_i \perp \{Y_i(0), Y_i(1)\}.$$ 
We assume that $\pi$ is bounded away from zero and one. 
Moreover, assume 
that the realized outcomes are 
$$Y_i = T_i Y_i(1) + (1-T_i)Y_i(0).$$
The (super-population) average treatment effect (ATE) is defined as
\begin{equation*}
    \tau := \mathbb{E}[Y(1) - Y(0)] \in \mathbb{R}^p.
\end{equation*}

\paragraph{Estimation and Inference of ATE.} Here, we review the standard approach to estimating $\tau$ and quantifying uncertainty when the dimension of the outcome, $p$,  does not increase at the same rate as $n$ and $p \ll n$. 
The standard approach to estimation of $\tau$ is to consider the {\em difference-in-means} estimator: 
   \begin{equation}\label{eq:dim}
       \hat{\tau}_{\text{DiM}} =  \frac{1}{n_t} \sum_{i=1}^n T_iY_i(1) - \frac{1}{n_c} \sum_{i=1}^n (1-T_i)Y_i(0),
   \end{equation}
   where $n_t = \sum_i T_i$ is the number of treated samples and $n_c = \sum_i (1-T_i)$ is the number of control samples.

Inference for $\hat{\tau}_{\text{DIM}}$ is obtained using a standard application of the central limit theorem.  Formally, we have:
   \begin{equation*}
       \sqrt{n}(\hat{\tau}_{\text{DiM}} - \tau) \xrightarrow[]{} N(0, \Sigma_{\text{DiM}})
   \end{equation*}
   where
   \begin{equation}
       \Sigma_{\text{DiM}} = \frac{1}{\pi}\text{Var}(Y(1)) + \frac{1}{1-\pi}\text{Var}(Y(0)), \label{eq:sigma}
   \end{equation}
   which can be consistently estimated as:
   \begin{align}
     \hat{\Sigma}_{\text{DiM}} &=  \frac{n}{n_t}\cdot\frac{1}{n_t} \sum_{i:T_i = 1} \left( Y_i(1) -  \frac{1}{n_t} \sum_{i:T_i = 1} Y_i(1)\right)\left( Y_i(1) -  \frac{1}{n_t} \sum_{i:T_i = 1} Y_i(1)\right)^{\intercal}  \\
     & + \frac{n}{n_c}\cdot \frac{1}{n_c} \sum_{i:T_i = 0} \left(Y_i(0) - \frac{1}{n_c} \sum_{i:T_i = 0} Y_i(0)\right)\left( Y_i(0) - \frac{1}{n_c} \sum_{i:T_i = 0} Y_i(0)\right)^{\intercal}. \label{eq:sigma-dim}
\end{align}

Inference can be made for each component $j$ of the outcome, $Y_{.,j}$, with the usual z-test and multiple testing. The p-value for testing the null hypothesis $H_{0j}: \tau_j = 0$ that there is no treatment effect on $Y_{.,j}$, is defined as
\begin{equation}\label{eq:individual-pvalue}
    P\left(Z \geq \sqrt{n}(\hat{\Sigma}_{\text{DiM}, jj})^{-1/2}|{\hat{\tau}_{\text{DiM}}}_j|\right) \cdot p,
\end{equation}
where $Z$ is a standard Gaussian random variable and $p$ is a multiple testing correction factor. This approach sets the foundation for the baseline approach presented in Section~\ref{sec:baseline}.

Inference can also be made for the group of outcomes. This involves the following statistic, referred to as the Hotelling $T^2$ statistic, which generalizes the usual $t$-test:
\begin{equation}\label{eq:hotelling}
    n  {\hat{\tau}_{\text{DiM}}}^{\intercal}(\hat{\Sigma}_{\text{DiM}})^{-1}{\hat{\tau}_{\text{DiM}}}.
\end{equation}
Under the null hypothesis $H_0: \tau = 0$ that there is no treatment effect in any dimension, the Hotelling $T^2$ statistic in \eqref{eq:hotelling} follows the chi-squared distribution with $p$ degrees of freedom. Thus, the p-value for testing the null hypothesis $H_0$ is defined as 
\begin{equation}\label{eq:group-pvalue}
     P\left(\chi^2(p) \geq n  {\hat{\tau}_{\text{DiM}}}^{\intercal}(\hat{\Sigma}_{\text{DiM}})^{-1}{\hat{\tau}_{\text{DiM}}} \right)
\end{equation}
where $\chi^2(p)$ is the chi-squared random variable with $p$ degrees of freedom. This approach is leveraged by our proposed approach of Section~\ref{sec:proposed}.

\paragraph{Our objective.} As discussed in the introduction, we are primarily interested in settings where the treatment effect may be {\em sparse} within the $p$-dimensional outcome.  Formally, throughout the remainder of the paper, we suppose that there exists a set of dimensions $S_{\tau}$ such that the sparsity index of $S_{\tau}$ is $|S_{\tau}| := s_{\tau} < p$ and
    \begin{align}
         \mathbb{E}[Y_j|T = 1, X] =  \mathbb{E}[Y_j|T = 0, X] \quad & \text{for } j \notin S_{\tau}, \\
         \mathbb{E}[Y_j|T = 1, X] \neq  \mathbb{E}[Y_j|T = 0, X] \quad & \text{for } j \in S_{\tau}. \label{eq:target_set}
    \end{align}

Our goal in this paper is to present a practical regression-based method for approximate recovery of $S_{\tau}$, together with valid inference on treatment effects localized to the identified subset. 
Next, we present a simple, commonly employed baseline approach for this purpose (first without, then with pre-treatment covariates); we subsequently compare our proposed approach to this baseline.

\section{A Baseline Approach}\label{sec:baseline}

In this section, we present a simple, commonly used baseline approach for approximate recovery of $S_\tau$, that serves as a reference point.

\subsection{Subset selection}
A data scientist may be inclined to perform the dimension-by-dimension testing, with multiple testing correction, of Section~\ref{sec:preliminaries} and then select all of the dimensions with significant p-values.  If this approach is used with a pre-defined sparsity index (or bound on the sparsity index) $s_\tau$, then it is equivalent to ranking outcomes by their treatment effect sizes, and selecting outcome dimensions with the largest effect sizes. This provides the basis of a commonly used baseline approach and is equivalent to considering the following objective:
$$\hat{S}_{\tau} = \{j | \hat{\beta}_j \neq 0\},$$
where 
\begin{equation}\label{eq:dim-ell0}
    \hat{\beta} = \arg \min_{\beta} \Bigg|\Bigg|\left(\frac{{\hat{\tau}_{\text{DiM}}}_j}{{\sqrt{\hat{\Sigma}_{\text{DiM}}}_{jj}}}\right)_{j=1}^p - \beta \text{ }\Bigg|\Bigg|_2^2 \quad \text{subject to } ||\beta||_0 \leq s_{\tau}.
\end{equation}
Here, $||\beta||_0 = \sum_{j=1}^{p} 1\{\beta_j \neq 0\}$ is the $\ell_0$ norm of $\beta$. That is, $\hat{S}_{\tau}$ is  simply the top $s_{\tau}$ outcomes based on the effect sizes of average treatment effects.

\subsection{Inference via sample-splitting} \label{sec:sample-splitting}
We are not only interested in identifying the dimensions along which there may exist a treatment effect, but also performing inference along the selected dimensions.  To accomplish this, we consider two sets of data: one for selection and one for inference.  As discussed in the introduction, our data may naturally be divided in clinical settings by a POC study on which we select dimensions and a larger clinical trial on which we estimate treatment effects.  Alternatively, we can deploy a sample-splitting approach on a single pool of data. Regardless, the above selection step is applied to the first data split.

Having identified a subset $\hat{S}_{\tau}$ on the first data split, inference can proceed as in Section \ref{sec:preliminaries} on the second data split, but restricted to the set $\hat{S}_{\tau}$.  We can specialize the definitions in \eqref{eq:dim} and \eqref{eq:sigma-dim} to the subset $\hat{S}_{\tau}$, and carry out inference using the second data split.  Although subset selection on the first split in this baseline approach was carried out using an approach inspired by \eqref{eq:individual-pvalue}, inference on the second split can proceed according to either \eqref{eq:individual-pvalue} or \eqref{eq:group-pvalue} defined with these $\hat{S}_{\tau}$-specific quantities. Note that in \eqref{eq:individual-pvalue}, we would still apply a correction factor for multiple testing, but it is significantly smaller, especially when $s_\tau=|\hat{S}_{\tau}|$ is much smaller than $p$.

\subsection{Incorporation of pre-treatment covariates}
In causal inference, pre-treatment covariates are often leveraged to reduce the variance of the treatment effect estimators (\cite{deng2013improving}, \cite{lin2013agnostic}, 
\cite{ying2023variancereduction}). 
As an example, the CUPED estimator \citep{deng2013improving} estimates $\tau$ as
\begin{equation}\label{eq:cuped}
    {\hat{\tau}_{\text{CUPED}}}_j =  \frac{1}{n_t} \sum_{i=1}^n T_i(Y_{ij}(1) - \hat{\theta}_j^{\intercal}(X_i- \bar{X}))- \frac{1}{n_c} \sum_{i=1}^n (1-T_i)(Y_{ij}(0) - \hat{\theta}_j^{\intercal}(X_i- \bar{X})),
\end{equation}
where $\hat{\theta}_j$ is the vector of OLS coefficients of a linear regression of $Y_{ij}$, the pooled outcomes of control and treatment groups, on pre-treatment covariates $X_i$, for $j = 1, \dots, p$. Note that $\bar{X}$ is the mean of covariates on the pooled data. 

Lin's estimator \citep{lin2013agnostic} improves the CUPED estimator by running separate regressions in the treated and control groups, with the agnostic property that it does not induce higher variance than the difference-in-means estimator without any model assumptions. Lin's estimator estimates $\tau$ as
\begin{equation}\label{eq:lin}
    {\hat{\tau}_{\text{Lin}}}_j = \frac{1}{n_t} \sum_{i=1}^n T_i(Y_{ij}(1) - {\hat{\theta}^1_j}^{\intercal}(X_i - \bar{X})) - \frac{1}{n_c} \sum_{i=1}^n (1-T_i)(Y_{ij}(0) - {\hat{\theta}^0_j}^{\intercal}(X_i - \bar{X})),
\end{equation}
where $\hat{\theta}^w_{j}$ ($w = 0$ or $1$) is the vector of OLS coefficients of a linear regression of $Y_{ij}(w)$ on the pre-treatment covariates $X_i$, for $j = 1, \dots, p$. 

Note that both the CUPED estimator and Lin's estimator can be written similarly to the difference-in-means estimators as
\begin{equation}\label{eq:adjusted-dim}
     \hat{\tau}_{\mbox{method}j} = \frac{1}{n_t} \sum_{i=1}^n T_i\Tilde{Y}_{ij}(1) - \frac{1}{n_c} \sum_{i=1}^n (1-T_i)\Tilde{Y}_{ij}(0),
\end{equation}
where for the CUPED estimator, $\Tilde{Y}_{ij}$ is defined as 
\begin{equation}\label{eq:Y-cuped}
    \Tilde{Y}_{ij} = Y_{ij} - \hat{\theta}_j^{\intercal}X_i
\end{equation}
and for Lin's estimator, $\Tilde{Y}_{ij}$ is defined as
\begin{equation}\label{eq:Y-lin}
    \Tilde{Y}_{ij} =  Y_{ij} - \frac{n_c}{n}{\hat{\theta}_j^1}^{\intercal}X_i - \frac{n_t}{n}{\hat{\theta}_j^0}^{\intercal}X_i.
\end{equation}

To adapt the baseline approach to use pre-treatment covariates, instead of the simple difference-in-means estimator, one can use covariate-adjusted difference-in-means estimators. For either $\text{method} \in \{\text{CUPED}, \text{Lin}\}$, we consider the following objective:
\begin{equation}\label{eq:dim-ell0-adjusted}
   \hat{\beta} = \arg \min_{\beta} \Bigg|\Bigg|\left(\frac{{\hat{\tau}_{\text{method}}}_j}{{\sqrt{\hat{\Sigma}_{\text{method}}}_{jj}}}\right)_{j=1}^p - \beta \text{ }\Bigg|\Bigg|_2^2 \quad \text{subject to } ||\beta||_0 \leq s_{\tau}.
\end{equation} 
The variance estimate is defined as \eqref{eq:sigma-dim} but with $\Tilde{Y}$ instead of $Y$. Then, again $\hat{S}_{\tau} = \{j | \hat{\beta}_j \neq 0\}$.

With these definitions, inference proceeds again as in Section \ref{sec:preliminaries} using the second split with covariate-adjusted $\Tilde{Y}$, but restricted to the set $\hat{S}_{\tau}$. 

It is possible to show under appropriate regularity and model assumptions that this baseline approach can consistently recover $S_{\tau}$.  (For example, assumptions similar to those we make in our subsequent theoretical development in Section \ref{sec:recovery} would suffice.)

However, this approach may be quite sample inefficient in recovering the appropriate subset in the low signal-to-noise-ratio (SNR) regimes. This is because the baseline approach carries out detection independently in each dimension, without leveraging structure jointly across the dimensions. We show later via  synthetic and semi-synthetic simulation results that the baseline approach exhibits a failure to select the right subset in the low SNR regime (see Figure \ref{fig:recovery-rate} and \ref{fig:recovery-rate-2} in Section \ref{sec:num-exp}). Notably, the examples discussed in Section~\ref{sec:introduction} generally fall in these low SNR regimes.  Motivated by this challenge, in the next section we take a sparse-linear-regression-based approach to subset selection, using the Lasso, which is known to generally perform well in low SNR regimes, cf.~\cite{hastie2020comparisons}.

\section{Our Proposed Approach:\\ Subset Selection via Sparse Regression}\label{sec:proposed}

Our approach starts from the following observation: Rather than considering each individual outcome dimension separately, they can be considered as a group as represented by the Hotelling's $T^2$ test statistic in \eqref{eq:hotelling}.   
This leads us to consider the following objective:
\begin{equation} \label{eq:power-ell0}
\hat{S}_{\tau} = \max_S {\hat{\tau}^S_{\text{DiM}}}^{\intercal}(\hat{\Sigma}_{\text{DiM}}^S)^{-1}{\hat{\tau}^S_{\text{DiM}}}  \quad \text{subject to } |S| \leq s_{\tau},
\end{equation}
where $\hat{\tau}^S_{\text{DiM}}$, $\hat{\Sigma}_{\text{DiM}}^S$ on the right-hand side are computed for each set $S \subset \{1,\dots,p\}$. 

In the next subsection, we show that, in fact, the preceding problem is equivalent to selecting $\hat{S}_{\tau} = \{j |\hat{\beta}_j \neq 0\}$ where $\hat{\beta}$ is estimated by
fitting a weighted linear regression with carefully chosen weights $W_i$ as
\begin{equation} \label{eq:wls-ell0}
    \hat{\beta} =  \arg\min_{\beta } \frac{1}{n}\sum_i W_i(T_i - (Y_i- \overline{Y})^{\intercal}{\beta})^2 \quad \text{subject to } ||\beta||_0 \leq s_{\tau}.
\end{equation}
Again, $||\beta||_0 = \sum_{i=1}^p 1\{\beta_i \neq 0\}$ is the $\ell_0$ norm of $\beta$. Note that we have treatment assignments as response variables and centered outcomes as explanatory variables. Here, $\bar{Y}$ is the mean of outcomes on the pooled data. 

The objective in \eqref{eq:power-ell0} represents a combinatorially hard problem in $p$, which we assume is large. Of note, the objective in \eqref{eq:wls-ell0} still presents computational complexity challenges as it does not yield a closed-form solution as did the baseline objective of \eqref{eq:dim-ell0}.  However, by reformulating \eqref{eq:power-ell0} as a weighted linear regression with subset selection \eqref{eq:wls-ell0}, we can employ various computationally efficient statistical tools that have been developed for the regression setting, such as forward subset selection, backward subset selection, or the Lasso. In the following, we focus on Lasso, which solves a convex relaxation of \eqref{eq:wls-ell0} by replacing the $\ell_0$ norm with the $\ell_1$ norm:
\begin{equation*}
    \hat{\beta} =  \arg\min_{\beta } \frac{1}{n}\sum_i W_i(T_i - (Y_i- \overline{Y})^{\intercal}{\beta})^2 \quad \text{subject to } ||\beta||_1 \leq t ,
\end{equation*}
where $||\beta||_1 = \sum_{i=1}^p |\beta_i|$ is the $\ell_1$ norm of $\beta$, and $t \geq 0$ is a tuning parameter. Due to convex duality, the above problem is equivalent to the more common penalized regression form 
\begin{equation}\label{eq:Lasso}
    \hat{\beta} =  \arg\min_{\beta } \frac{1}{n}\sum_i W_i(T_i - (Y_i- \overline{Y})^{\intercal}{\beta})^2  + 2\lambda ||\beta||_1,
\end{equation}
where $\lambda \geq 0$ is a tuning parameter. Again,
$\hat{S}_{\tau} = \{j| \hat{\beta}_j \neq 0\}.$ 

Having identified a subset $\hat{S}_{\tau}$ on the first data split, inference proceeds exactly as in Section~\ref{sec:sample-splitting}.

The pre-treatment covariate adjusted version of our proposed method is straightforward: we directly include pre-treatment covariates in the regression.  In particular, for the Lasso, instead of \eqref{eq:Lasso}, we fit a weighted Lasso of treatment assignments on centered outcomes and centered covariates as follows:

\begin{equation}\label{eq:Lasso_adjusted}
    (\hat{\beta}, \hat{\alpha}) =  \arg\min_{\beta, \alpha }\frac{1}{n}\sum_i W_i(T_i - (Y_i- \overline{Y})^{\intercal}{\beta} - (X_i- \overline{X})^{\intercal}{\alpha})^2  + 2\lambda ||\beta||_1.
\end{equation}
Note that
\begin{align*}
    &\arg\min_{\beta} \min_{\alpha}\frac{1}{n}\sum_i W_i(T_i - (Y_i- \overline{Y})^{\intercal}{\beta} - (X_i- \overline{X})^{\intercal}{\alpha})^2  \\
    & =  \arg\min_{\beta}\frac{1}{n}\sum_i W_i (T_i - (\Tilde{Y}- \Bar{\Tilde{Y}})^{\intercal} \beta)^2, 
\end{align*}
where 
\begin{align}\label{eq:adjustY_joint}
    \Tilde{Y}_{ij} = Y_{ij} - X_i^{\intercal} \left[\left(\frac{1}{n}\sum_{i=1}^n W_i (X_i - \Bar{X})(X_i - \Bar{X})^{\intercal}\right)^{-1} \left(\frac{1}{n}\sum_{i=1}^n W_i (X_i - \Bar{X})(Y_{ij} - \Bar{Y}_j)\right)\right].
\end{align}
Therefore, our approach to include pre-treatment covariates closely resembles the adjustment of the CUPED estimator or Lin's estimator.  In particular, we adjust the outcomes $Y$ with pre-treatment covariates using weighted linear regression.

Having identified $\hat{S}_{\tau}$, using the second split, we compute the covariate-adjusted difference-in-means estimator such as the CUPED estimator or Lin's estimator on $\hat{S}_{\tau}$. The inference proceeds again as in Section \ref{sec:preliminaries} with covariate-adjusted $\Tilde{Y}$ restricted to the set $\hat{S}_{\tau}$.

Our method shares a close relationship with the projection test \citep{lopes2011powerful, huang2015projection, liu2022msptesting} in high-dimensional mean testing, as both involve finding sparse $\beta$ using a quadratic objective function with $\ell_1$ penalization. We focus on causal inference, and in particular as a result, we also consider pre-treatment covariates. Moreover, we employ a linear regression specification, making it easier for practitioners to implement our method with tools like \texttt{glmnet} in R.

Algorithm~\ref{algo:single-split} provides a summary of the proposed method given two sets of data. As mentioned before, our data may naturally be divided in clinical settings by a POC study and a larger clinical trial after. Alternatively, we can employ a sample-splitting approach with a single pool of data. For a sample-splitting approach, the results are often sensitive to the arbitrary choice of a one-time random split of the data \citep{meinshausen2009pvalues}. Therefore, it is recommended to employ multiple splits and aggregate p-values from each split. In Appendix \ref{sec:mss}, we present how we apply multi-sample splitting results from \cite{meinshausen2009pvalues} to our proposed method when two data sets are obtained through sample-splitting.

\begin{algorithm}
\DontPrintSemicolon 
   \vspace*{0.2cm}
  \KwInput{$\mathcal{D}_1 = \{(T_i, X_i, Y_i)\}_{i=1}^{n_1}$ and $\mathcal{D}_2 = \{(T_i, X_i, Y_i)\}_{i=1}^{n_2}$}
  \KwOutput{$\hat{S}_{\tau}$ and the estimated treatment effects on $\hat{S}_{\tau}$}
  \vspace*{0.2cm}
  Using the first split $\mathcal{D}_1$, estimate $$\hat{S}_{\tau} = \{j| \hat{\beta}_j \neq 0\}$$ by fitting a weighted Lasso regression of $T$ on centered $Y$ and $X$ as in \eqref{eq:Lasso_adjusted}.\\
  \vspace*{0.2cm}
  Using the second split $\mathcal{D}_2$, estimate the treatment effect $\hat{\tau}$ such that for $j \in \hat{S}_{\tau}$,
  \begin{equation*}
     \hat{\tau}_j = \frac{1}{\sum_{i\in \mathcal{D}_2}T_i} \sum_{i\in \mathcal{D}_2} T_i\Tilde{Y}_{ij}(1) - \frac{1}{\sum_{ i\in \mathcal{D}_2}(1-T_i)} \sum_{i\in \mathcal{D}_2} (1-T_i)\Tilde{Y}_{ij}(0),
  \end{equation*} 
  where $\Tilde{Y}_{ij}$ is defined as in \eqref{eq:Y-cuped} or \eqref{eq:Y-lin}. Its variance estimate $\hat{\Sigma}$ can be obtained using \eqref{eq:sigma-dim} with $\Tilde{Y}$ restricted to the set $\hat{S}_{\tau}$.
  \vspace*{0.2cm}
\caption{Subset Selection via Sparse Regression (Single-Split)}\label{algo:single-split}
\end{algorithm}

In the remainder of the section, we first discuss the reduction of \eqref{eq:power-ell0} to a weighted least squares regression; we then present a result that provides conditions under which we recover $S_{\tau}$.  

\subsection{Reduction to Weighted Linear Regression}
\label{sec:wls}

In the following, we show that the weighted least-squares problem in \eqref{eq:wls-ell0} is equivalent to maximization of the Hotelling $T^2$ statistic in \eqref{eq:power-ell0}, which motivates the use of sparse linear regression in our approach.

Consider the following weighted least squares regression:
\begin{equation}\label{eq:wls}
    \hat{\beta} =  \arg\min_{\beta } \frac{1}{n}\sum_iW_i(T_i - (Y_i^S- \overline{Y^S})^{\intercal}{\beta})^2,
\end{equation}
where $Y_i^S$ is an outcome vector restricted to the given set $S$ as
$Y_i^S := (Y_{i,j} |j \in S)^{\intercal}.$
We set the weights $W_i$ so that the least-squares problem becomes equivalent to a \emph{signal strength maximization} problem, where the signal strength is defined using Hotelling $T^2$ statistic in \eqref{eq:hotelling}.

To start, consider the unweighted setting. Here, the residual sum of squares (RSS) of the linear regression is
\begin{align}
    \frac{1}{n} \sum_i (T_i - (Y^{S}_i- \overline{Y^S})^{\intercal}\hat{\beta})^2 &= C_1 - C_2 \cdot {\hat{\tau}_{\text{DiM}}^S}^{\intercal}   \left(\hat{\Sigma}^S + C_3 \cdot \hat{\tau}_{\text{DiM}}^S{\hat{\tau}_{\text{DiM}}^S}^{\intercal}\right)^{-1} \hat{\tau}_{\text{DiM}}^S \\
    &= C_1 - C_2 \cdot \frac{{\hat{\tau}_{\text{DiM}}^S}^{\intercal}(\hat{\Sigma}^S)^{-1}\hat{\tau}_{\text{DiM}}^S}{1 + C_3  \cdot {\hat{\tau}_{\text{DiM}}^S}^{\intercal}(\hat{{\Sigma}}^S)^{-1}\hat{\tau}_{\text{DiM}}^S} \label{eq:rss-power}
\end{align}
by applying the Sherman–Morrison formula. Note that $C_1, C_2, C_3 >0$. Details can be found in the Appendix, Section \ref{sec: calculations}. The $ \hat{\tau}_{\text{DiM}}^S$ is defined as in \eqref{eq:dim} (restricted to the subset $S$) and $\hat{\Sigma}^S$ is defined as
\begin{align}
     \hat{\Sigma}^S &=  \frac{n}{n_c}\cdot\frac{1}{n_t} \sum_i T_i( Y^S_i(1) -  \overline{Y^S(1)})( Y^S_i(1) -  \overline{Y^S(1)})^{\intercal}  \\
     & + \frac{n}{n_t}\cdot \frac{1}{n_c} \sum_i (1-T_i)( Y^S_i(0) - \overline{ Y^S(0)})( Y^S_i(0) - \overline{Y^S(0)})^{\intercal}. \label{eq:sigmahat_prev} 
\end{align}
We want to correct $\hat{\Sigma}^S$ by introducing weights $W_i$ so that $\hat{\Sigma}^S$ becomes $\hat{\Sigma}_{\text{DiM}}^S$.

We design the following weights:
$$
W_i = \frac{1}{\hat{\pi}^2}T_i +  \frac{1}{(1-\hat{\pi})^2}(1-T_i),$$ where $\hat{\pi} = \sum_i T_i /n$. 
Under these weights, the RSS of the weighted linear regression is now
\begin{align}
     \sum_i W_i(T_i - (Y^S_i- \overline{Y^S})^{\intercal}\hat{\beta})^2 = C_1 - C_2 \cdot \frac{{\hat{\tau}_{\text{DiM}}^S}^{\intercal}(\hat{\Sigma}_{\text{DiM}}^S)^{-1}\hat{\tau}_{\text{DiM}}^S}{1 + \left(\frac{n_c}{n_t}+\frac{n_t}{n_c} - 1\right) \cdot {\hat{\tau}_{\text{DiM}}^S}^{\intercal}(\hat{\Sigma}_{\text{DiM}}^S)^{-1}\hat{\tau}_{\text{DiM}}^S},\label{eq:wls-rss-power}
 \end{align}
where $C_1, C_2>0$. Details can be found in the Appendix, Section \ref{sec: calculations}. 

In \eqref{eq:wls-rss-power}, $C_1, C_2, C_3$ do not depend on the given set $S$ but $\hat{\tau}_{\text{DiM}}^S$ and $\hat{\Sigma}_{\text{DiM}}^S$ do.
This indicates that for any $s \leq p$, for a subset of outcomes $S$ such that $|S| = s$, we have
\begin{align}\label{eq:signal-maximization}
    \min_{S : |S| = s} \text{  }\text{RSS of weighted linear regression with }Y^S = \max_{S : |S| = s} \text{  }{\hat{\tau}_{\text{DiM}}^S}^{\intercal}(\hat{\Sigma}_{\text{DiM}}^S)^{-1}\hat{\tau}_{\text{DiM}}^S.
\end{align}
On the right-hand side, ${\hat{\tau}_{\text{DiM}}^S}$ and $\hat{\Sigma}_{\text{DiM}}^S$ are obtained for each $S$. Note that the right-hand side is the Hotelling $T^2$ statistic in \eqref{eq:hotelling}. This means that with our chosen weights, the weighted least-squares problem in \eqref{eq:wls-ell0} becomes equivalent to maximization of the Hotelling $T^2$ statistic as in \eqref{eq:power-ell0}, as desired.

\subsection{Recovery Results for the Lasso}
\label{sec:recovery}

In this section, we provide theoretical results for our approach using the Lasso.  In particular, we state the assumptions under which our proposed method can consistently recover $S_{\tau}$, and provide the error bound for any optimal solution of \eqref{eq:Lasso_adjusted}.  

Recall that $\hat{\beta}$ is defined as the solution of the empirical risk:
\begin{equation}\label{eq:emp_risk}
    (\hat{\beta}, \hat{\alpha}) = \arg\min_{\beta \in \mathbb{R}^{p}, \alpha \in \mathbb{R}^m}\frac{1}{n} \sum_i W_i( T_i -  ( Y_i - \overline{Y})^{\intercal} \beta -  (X_i - \overline{X})^{\intercal} \alpha)^2 + 2\lambda_n ||\beta||_1,
\end{equation}
where $\lambda_n > 0$ is a user-defined regularization penalty. 

Define the optimal solution of the population risk, $\beta^*$, as 
\begin{equation}\label{eq:pop_risk}
    \beta^* = \arg\min_{\beta \in \mathbb{R}^{p}} \min_{\alpha \in \mathbb{R}^m}\mathbb{E}\Bigg[ \frac{1}{n} \sum_i W_i^*( T_i -  ( Y_i - \mathbb{E}[Y])^{\intercal} \beta -  (X_i - \mathbb{E}[X])^{\intercal} \alpha)^2 \Bigg],
\end{equation}
where $W_i^* = \frac{1}{\pi^2}T_i + \frac{1}{(1-\pi)^2}(1-T_i)$. Then, we have
\begin{equation*}
    \beta^* = \arg\min_{\beta \in \mathbb{R}^{p}} \mathbb{E}\Bigg[ \frac{1}{n} \sum_i W_i^*( T_i -  ( Z_i - \mathbb{E}[Z])^{\intercal} \beta)^2 \Bigg],
\end{equation*}
where $Z_i$ is the (population-level) covariate-adjusted outcome defined as
\begin{equation*}
    Z_i := Y_i - (1-\pi)\cdot \left[\text{Var}(X)^{-1}\text{Cov}(X, Y(1))\right]^{\intercal}X_i - \pi \cdot \left[\text{Var}(X)^{-1}\text{Cov}(X, Y(0))\right]^{\intercal}X_i.
\end{equation*}
Note that this adjustment of outcomes $Y$ on pre-treatment covariates $X$ is very similar to pre-treatment covariates adjustments with linear regressions in Lin's estimator \citep{lin2013agnostic}. If pre-treatment covariates are not available, we simply set $Z = Y$.

The closed form of $\beta^*$ can be verified to be:
\begin{equation}\label{eq:beta-star}
    \beta^* = \frac{1-\pi}{\pi} \left(\Sigma_Z + \left(\frac{1-\pi}{\pi} + \frac{\pi}{1-\pi} - 1\right) \tau\tau^{\intercal}\right)^{-1}\tau.
\end{equation} 
where
\begin{align}
    \tau &:= \mathbb{E}[Z(1) - Z(0)] = \mathbb{E}[Y(1) - Y(0)], \label{eq:tau_Z}\\
    \Sigma_Z &:= \frac{1}{\pi} \text{Var}(Z(1)) + \frac{1}{1-\pi}\text{Var}(Z(0)) \label{eq:sigma_Z}.
\end{align}
By the Sherman-Morrison formula, we have
\begin{equation*}
    \beta^* \propto \Sigma_Z^{-1}\tau.
\end{equation*}

In the following, we provide an error bound for $||\hat{\beta} - \beta^*||$, and show that $\hat{S}_{\tau} = \{j|\hat{\beta}_j\neq0\}$
recovers the set $S^* := \{j|\beta^*_j \neq 0\}$ asymptotically. However, note that even if the treatment effect is sparse---i.e., even if $|S_\tau| < p$ where $S_{\tau}= \{j| \tau_j \neq 0\}$ in \eqref{eq:target_set}---it does not follow that $\beta^*$ is sparse.  To ensure $\beta^*$ is sparse, we additionally require 
the inverse of $\Sigma_Z$ to be sparse. Below, we provide examples of structural causal models \citep{bollen1989,pearl2009} where the inverse of $\Sigma_Z$ is sparse, resulting in a sparse $\beta^*$, with the set $S^*$ containing the target set $S_{\tau}$. We formalize this as the following sparsity assumption on $\beta^*$ for future reference. We also provide simulation results when the sparsity assumption is slightly violated in the Appendix \ref{sec:add_sims}.

\begin{assumption}[Sparsity]\label{ass-sparsity} $\beta^*$ is sparse with only $s^*$ nonzero elements. 
\end{assumption}

\begin{example} Consider the linear structural causal model:
\begin{align*}
    Y &= \bm{\alpha}_0 + \Theta_0X + T\cdot(\bm{\alpha}_1 + \Theta_1X) +  \epsilon,
\end{align*}
where $\bm{\alpha}_0, \bm{\alpha}_1 \in \mathbb{R}^p$, $\Theta_0, \Theta_1 \in \mathbb{R}^{p\times m}$, and $ X \perp \epsilon$. Suppose that the treatment effect is sparse; there exists $S_{\tau}$ such that for $j \notin S_{\tau}$, we have $\bm{\alpha}_{1j} = 0$ and the $j$-th row $\Theta_{1}^j = \bm{0}$. 

Assume that $\epsilon_{1}, \dots, \epsilon_{p}$ are mean zero and uncorrelated with each other. Then, we have $S_\tau = S^*$.
\end{example}

\begin{example}
In this example we consider the same linear structural causal model as in the previous example, but instead of assuming that $\epsilon_{1}, \dots, \epsilon_{p}$ are uncorrelated with each other, we instead assume a cluster structure: there exist clusters $C_1, \dots, C_c$ such that if $j \in C_i$ and $j' \in C_{i'}$ with $i \neq i'$, then $\epsilon_j$ and $\epsilon_{j'}$ are uncorrelated. Then, when $S_{\tau} \subseteq C_1$, we have
 \begin{align*}
        S_{\tau}  \subseteq S^* \subseteq C_1.
    \end{align*}
With a sparse cluster structure, $s_{\tau} \leq s^* \leq |C_1| \ll p$.
\end{example}

\begin{example}
Continue to assume the same linear structural causal model, but now assume that $\epsilon$ follows a multivariate normal distribution and ``most" variables are conditionally independent of each other given the remaining variables. For example, we could consider an autoregressive model (AR($k$)) for time-series data. Let
$$
S_{\tau} \subseteq C :=\{j \text{ }|\text{ } \exists i\in S_{\tau} \text{ such that } \epsilon_i \text{ and } \epsilon_j \text{ are not conditionally independent.}\}
$$
Then, the inverse covariance matrix $\Sigma_Z$ is sparse. Specifically, if $i \in S_{\tau}$ and $j \notin C$, $(\Sigma_Z^{-1})_{i,j} = 0$. Therefore, we have 
\begin{equation*}
    S_{\tau} \subseteq S^* = C. 
\end{equation*}
With a sparsely connected graph structure for $\epsilon$, $s^* \ll p$.
\end{example}

For our theoretical results, we assume sub-Gaussian designs for $X$, $Y(0)$, $Y(1)$, and thus for $Z(0), Z(1)$ as well. Then, we make additional assumptions about their moments and the invertibility of the covariance matrix of $X$.

\begin{assumption}[Sub-Gaussian Designs]\label{ass:sub-gaussianity}
$X_1, \dots, X_n$ are mutually independent and identically distributed sub-Gaussian vectors with mean $\mu_X$ and covariance $\Sigma_X$.  Moreover, $Y_1(0), \dots, Y_n(0)$ and $Y_1(1), \dots, Y_n(1)$ are mutually independent and identically distributed sub-Gaussian vectors with mean $\mu^0_Y$ and $\mu^1_Y$ respectively and covariance $\Sigma^0_Y$ and $\Sigma^1_Y$ respectively. Additionally, $Z_1(0), \dots, Z_n(0)$ and $Z_1(1), \dots, Z_n(1)$ are also mutually independent and identically distributed sub-Gaussian vectors. Let the mean  be $\mu^0_Z$ and $\mu^1_Z$ respectively and the covariance be $\Sigma^0_Z$ and $\Sigma^1_Z$ respectively. More specifically, 
each variable $(Z(0))_j/\sqrt{(\Sigma^0_Z)_{jj}}$ is sub-Gaussian with parameter $\sigma_Z^0$ and $(Z(1))_j/\sqrt{(\Sigma^1_Z)_{jj}}$ is sub-Gaussian with parameter $\sigma_Z^1$ for $j = 1, \dots, p$. 
\end{assumption} 

\begin{assumption}[Moments]\label{ass:bound} The first and second moments of $X, Y(0), Y(1), Z(0), Z(1)$ are bounded. 
\end{assumption}

\begin{assumption}[Invertibility of $\Sigma_X$]\label{ass:eigen} The minimum eigenvalue of $\Sigma_X$ is bounded away from zero. 
\end{assumption}

A crucial assumption in the proof of the oracle inequalities for Lasso is the {\em restricted eigenvalue (RE)} condition introduced by \cite{bickel2009Lasso} for the covariance matrix. When $p > n$, the sample covariance matrix is not positive definite. RE conditions imply a kind of restricted positive definiteness. Let $J(\beta) = \{j \in \{1, \dots, p\}: \beta_j \neq 0\}$. For a vector $\bm{\delta}\in \mathbb{R}^p$ and a subset $J \subset \{1, \dots, p\}$, we denote by $\bm{\delta}_J$ the vector in $\mathbb{R}^p$ that has the same coordinates as $\bm{\delta}$ on $J$ and zero coordinates on the complement $J^c$ of $J$.
We say that $\Sigma$ satisfies the RE$(s, c_0, \Sigma)$ condition with parameter $\kappa(s, c_0, \Sigma)$ if 
\begin{equation}\label{eq:RE}
     \kappa(s, c_0, \Sigma) = \min_{\substack{J_0 \subseteq \{1, \dots, p\} \\ |J_0| \leq s}} \min_{\substack{\bm{\delta} \in \mathbb{R}^{p}, \bm{\delta}\neq 0, \\ ||\bm{\delta}_{J_0^c}|| \leq c_0 ||\bm{\delta}_{J_0}||_1}} \frac{\bm{\delta}^{\intercal} \Sigma \bm{\delta}}{||\bm{\delta}_{J_0}||_2^2} > 0. 
\end{equation}
Several sufficient conditions for RE conditions can be found in \cite{bickel2009Lasso}.

In Assumption \ref{ass:re}, we assume appropriate RE conditions as well as additional sparse eigenvalue conditions for population covariance matrices of $Z$, $\Sigma_Z^0$ and $\Sigma_Z^1$.  Essentially, this suggests that the covariance matrix of outcomes, after (population-level) linear covariates adjustment, satisfies some \emph{restricted} positive definiteness.

\begin{assumption}[RE condition]\label{ass:re} 
    $\Sigma_Z^0$ satisfies RE $(s^*, 3, \Sigma_Z^0)$ condition with parameter $\kappa(s^*, 3,\Sigma_Z^0)$ and $\Sigma_Z^1$ satisfies RE $(s^*, 3, \Sigma_Z^1)$ condition with parameter $\kappa(s^*, 3, \Sigma_Z^1)$. Furthermore, the maximum $s^*$-sparse eigenvalues for both groups are bounded, which means that for some constant $\rho>0$,
    \begin{equation*}
     \max_{\substack{\bm{||\delta}||_2=1,  ||\bm{\delta}||_0 \leq s^*}}\bm{\delta}^{\intercal}\Sigma_Z^0 \bm{\delta},  \quad \max_{\substack{||\bm{\delta}||_2=1, ||\bm{\delta}||_0 \leq s^*}}\bm{\delta}^{\intercal}\Sigma_Z^1 \bm{\delta} \quad \leq \rho.
    \end{equation*}
\end{assumption}

Finally, we assume that the optimal solution $\beta^*$ is well-behaved, with $||\beta^*||_1$ bounded. 
\begin{assumption}[Bounded $||\beta^*||_1$]
    There exists $M > 0$ such that $||\beta^*||_1 \leq M$.
\end{assumption}

In the following theorem, we first establish the error bound for $||\hat{\beta} - \beta^*||_1$, and show that we can recover $S^*$ consistently with the hard-threshold Lasso estimator. The proof largely follows the proof of Theorem 7.2 in \cite{bickel2009Lasso}, and can be found in the Appendix, Section \ref{sec:proofs}. Note that we allow the dimension of outcome representations to be $p \gg n$, but we assume that the dimension of pre-treatment covariates $m$ is fixed.

\begin{theorem}[Rate and Recovery]\label{main-theorem}
    Suppose assumptions stated above hold and 
    $\min (n_t/n, n_c/n) \geq r > 0$. 
    Consider $\hat{\beta}$ defined in \eqref{eq:Lasso_adjusted} with 
    $\lambda_n \geq A \sqrt{\log p /n}$
    for some sufficiently large $A >0$. 
    Then, for sufficiently large $n \gg s^*\log p$, we have
    \begin{align*}
        ||\hat{\beta} - \beta^*||_1 & \leq c_3 \text{  } s^*\lambda_n,
    \end{align*}
    with probability at least $1 - c_1 \exp{\left(-c_2  n \lambda_n^2\right)}$ for some constants $c_1, c_2, c_3 > 0$. Further, if
    \begin{equation*}
     \min_{j \in S^*} |\beta^*_j| \gg c_4  s^*\lambda_n \quad \text{as } n \xrightarrow[]{}\infty,
 \end{equation*}
    for some constant $c_4 > 0$
     and $\hat{S^*}$ is defined using the hard-threshold Lasso estimator as
     $$
     \hat{S}^* = \{ j \text{ }|\text{ } |\hat{\beta}_j| > c_4   s^* \lambda_n\},
     $$
     then we have a consistent recovery of $S^*$ as $\mathbb{P}(\hat{S}^* = S^*) \xrightarrow[]{}1  \text{  } \text{as } n \xrightarrow[]{} \infty. $
\end{theorem}

 \begin{remark}
     Consider the classical setting, where the number of samples $n$ goes to infinity, but 
     the sparsity index $s^*$ do not depend on $n$. One could choose $\lambda_n^2 = 1/n^{1-\delta}$ for some $\delta > 0$. Then with probability greater than $1-\tilde{c}_1\exp(-\tilde{c}_2 n^{\delta})$, which converges to 1 as $n \xrightarrow[]{} \infty$, we have an error bound for $||\hat{\beta} - \beta^*||_1$ that converges to 0.
\end{remark}

\section{Numerical Experiments}\label{sec:num-exp}

\subsection{Simulations}\label{sec:simulations}

We compare the ranking performance and power of the baseline and the proposed methods with pre-treatment covariate adjustment, as presented in Section \ref{sec:proposed}, through a simulation study. In scenarios where the data generation process involves independent outcomes and a constant treatment effect without pre-treatment covariates, both the baseline and proposed approaches demonstrate similar performance. Additional details can be found in the Appendix \ref{sec:add_sims}. Therefore, we focus on scenarios where pre-treatment covariates explain some of the variance in outcomes. 

We generate $n$ samples in the following manner:
\begin{align}
    &Y_{ij} = X_i^{\intercal}\beta_j + T_i \cdot (\alpha + X_i^{\intercal} \delta_j)  \cdot I(j \leq s_{\tau}) + \epsilon_{ij},\\
     &X_i \stackrel{i.i.d.}{\sim} N(0, I_{m \times m}), \text{  }
     T_i \stackrel{i.i.d.}{\sim} \text{Bernoulli}(\pi), \text{  }
    \epsilon_{ij} \stackrel{i.i.d.}{\sim}  N(0, 1),
\end{align}
for $i = 1, \dots, n$ and $j = 1, \dots, p$, where $p$ is the outcome dimension. The model parameters are first randomly sampled as
\begin{align*}
    \beta_j = (U_{j1}, \dots, U_{jm}) \quad &\text{where } U_{j1}, \dots, U_{jm} \stackrel{i.i.d.}{\sim} \text{Unif}(-1, 1) \\
    \delta_{j} = (U'_{j1}, \dots, U'_{jm}) \quad &\text{where } U'_{j1}, \dots, U'_{jm} \stackrel{i.i.d.}{\sim} \text{Unif}(0, 1)
\end{align*}
for $j = 1, \dots, p$.  
Note that there are non-zero treatment effects on $S_{\tau} = \{Y_1, \dots, Y_{s_{\tau}}\}$, where $s_{\tau}$ is a pre-specified sparsity index in the simulation process.
We set the outcome dimension $p = 500$, the number of pre-treatment covariates $m = 50$, and the sparsity index $s_{\tau} = 5$.

We evaluate the performance of the proposed method relative to the baseline method, both with adjustment for pre-treatment covariates, based on how well the ranking recovers $S_{\tau}$. The ranking is obtained as follows: For the baseline method, individual outcome dimensions are ranked based on their treatment effect sizes. For the proposed method, we consider regularization paths for a fine grid of $\lambda$ values and track a corresponding sequence of active sets (the set with non-zero coefficients) of increasing size. The \emph{recovery rate} is defined as follows: For a given $s$, denote $S(s)$ as the selected subset of outcome dimensions of size $s$ obtained from applying each method with pre-treatment covariates adjustment. Then, for each method, the recovery rate of $S_{\tau}$ is defined as $|S(s) \cap S_{\tau}|/s$ for $1 \leq s \leq s_{\tau}$.

Additionally, we evaluate statistical power of both methods. A second data set with a sample size of 500 was generated from the same data-generating process. For each selected subset of size $s$ from the first data set with a sample size of $n$, we evaluate power on the second data set for each method.

\begin{figure}
    \centering
    \includegraphics[scale = 0.8]{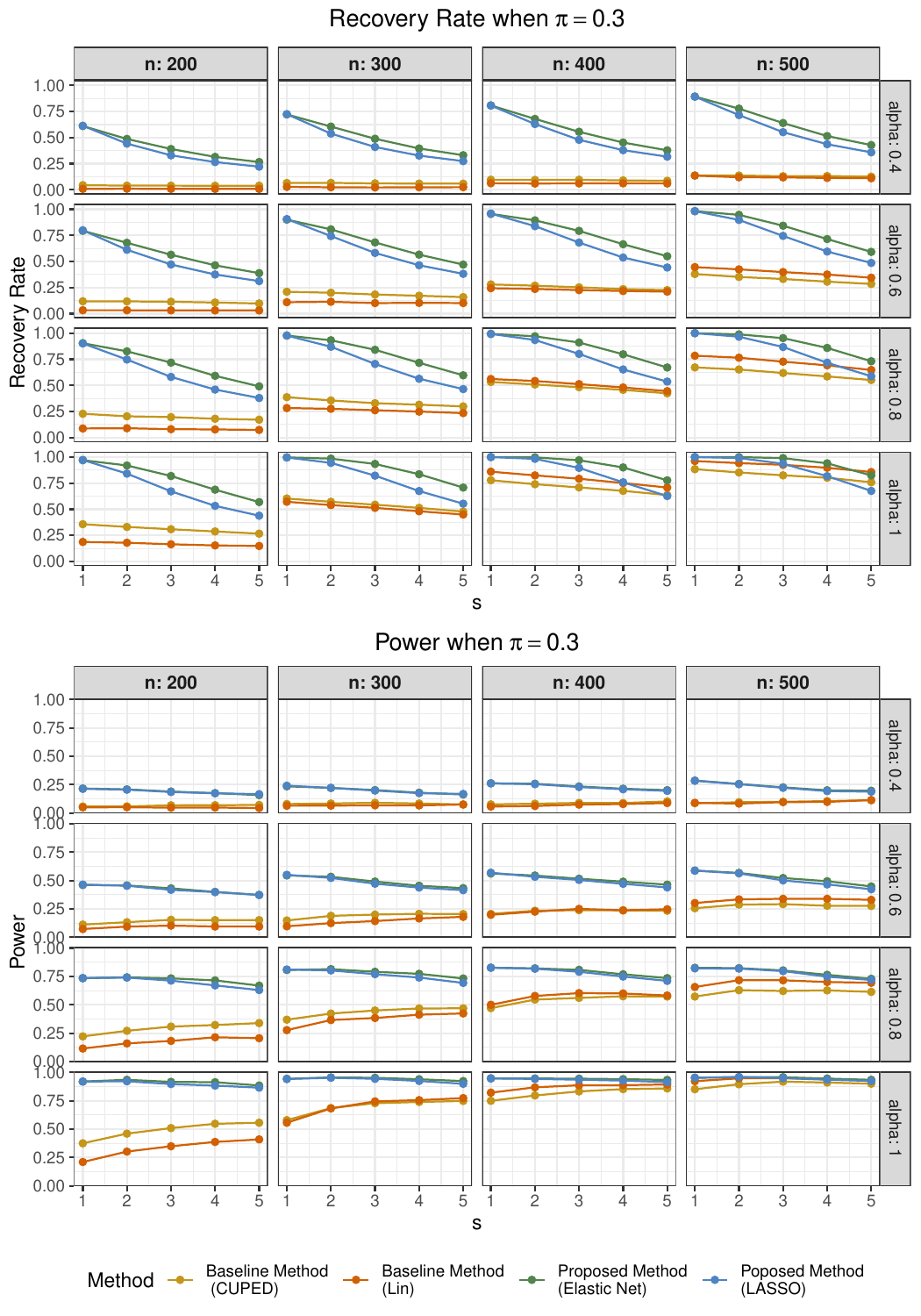}
    \caption{Recovery rates and powers averaged over 1000 replicates for the baseline and proposed approaches with pre-treatment covariates adjustment for $n = 200, 300, 400, 500$ and $\alpha = 0.4, 0.6, 0.8, 1$ for $\pi = 0.3$.}
     \label{fig:recovery-rate}
\end{figure}

\begin{figure}
    \centering
    \includegraphics[scale = 0.8]{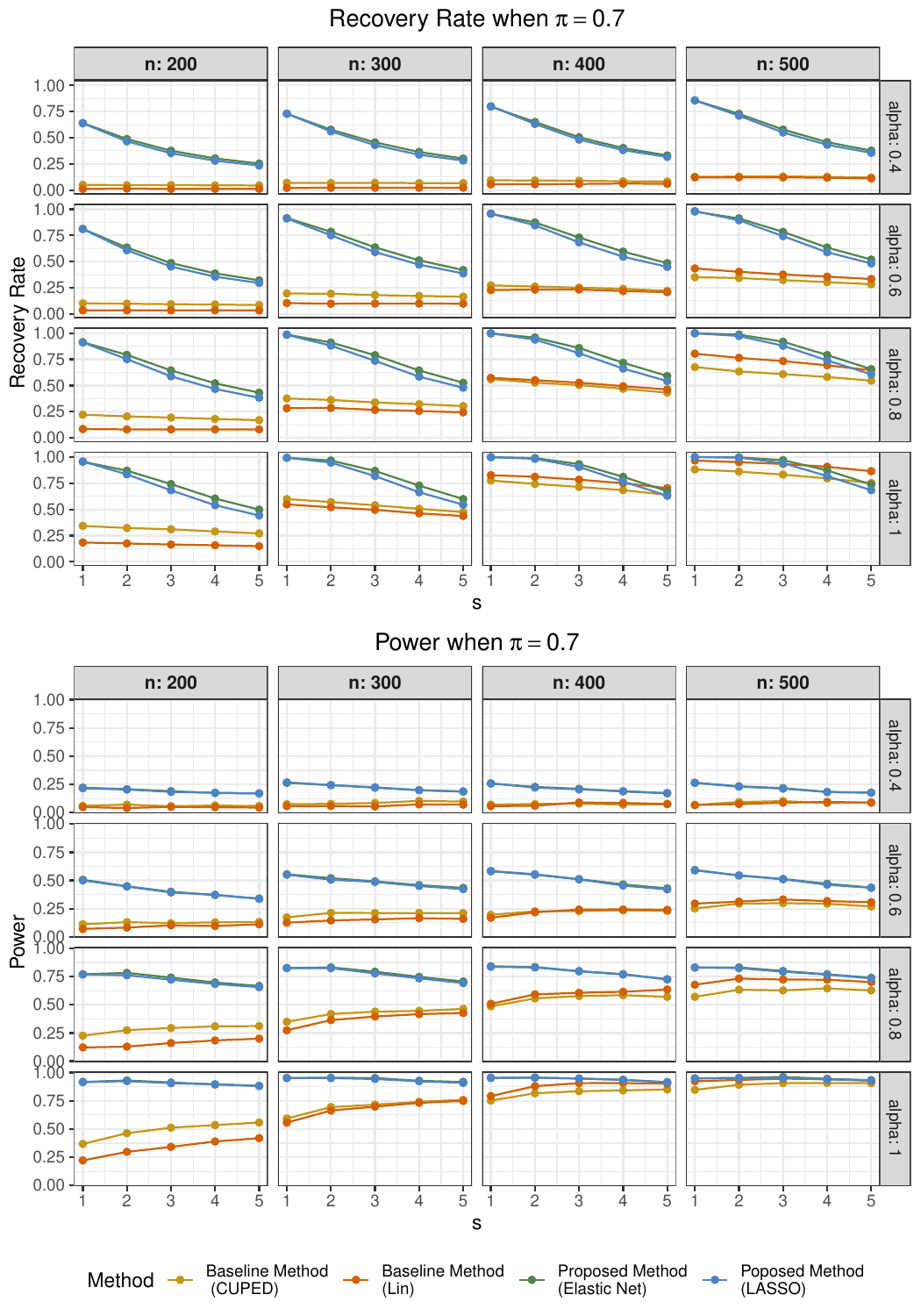}
    \caption{Recovery rates and powers averaged over 1000 replicates for the baseline and proposed approaches with pre-treatment covariates adjustment for $n = 200, 300, 400, 500$ and $\alpha = 0.4, 0.6, 0.8, 1$ for $\pi = 0.7$.}
     \label{fig:recovery-rate-2}
\end{figure}

The recovery rates and powers comparing the proposed to baseline approaches (with pre-treatment covariates adjustment) for different sample sizes ($n$), probabilities of being treated ($\pi$), and magnitudes of treatment effects ($\alpha$) are given in Figure \ref{fig:recovery-rate} and \ref{fig:recovery-rate-2}. Note that in our simulations, we use both the Lasso penalty and the elastic net penalty \citep{zou2005regularization}. The elastic net is known for encouraging a grouping effect, where correlated predictors tend to be included or excluded together. 
We see that our proposed method outperforms baseline methods in most scenarios, especially when sample sizes are small and signals are weak, with the elastic net achieving even better performance than the Lasso. When sample sizes and signals are large, the proposed method tends to identify a subset of $S_{\tau}$ more quickly than baseline methods, but performs slightly worse in discovering the full $S_{\tau}$. 

\subsection{Semi-Synthetic Data}
\label{sec:semisynth}

In this section, we demonstrate the performance of our method using semi-synthetic data motivated from the time series data example introduced in Section~\ref{sec:introduction}. 
Following the protocol discussed in Section~\ref{sec:introduction}, we curate a dataset with $n=5,000$ samples of the average 24-hour period for a week of real CGM data before and after a potential treatment. Specifically, we randomly sampled 10,000 two-week CGM traces from the CGM time-series data of 149 patients collected from the ongoing 4T study at Stanford's Lucille Packard Children's Hospital. While these traces are not fully independent—since multiple traces can come from the same patient—we assume independence for our semi-synthetic analysis. From the 10,000 traces, we retained the 5,000 with the fewest missing values.

For each sample $i$, let the transformed glucose traces be $\text{glucose}_i \in \mathbb{R}^{288\times 2 = 576}$, where 288 represents 24 hours of data recorded at 5 minute intervals averaged over the 7 days in a given week and we concatenate this 24-hour period average representation for Week 1 and Week 2. An example of such a transformed CGM trace is illustrated in Figure \ref{fig:cgm-example}.
Generalizing the procedure of Section~\ref{sec:introduction}, we generate semi-synthetic data by simulating a treatment effect that occurs within a random two-hour time interval of the day. The two-hour time interval is randomly selected across the 1000 repeats of the procedure. For example, the treatment effect might occur following lunch, between 12pm and 2pm. Given a randomly selected two-hour interval $I$ in which the treatment effect may manifest, a synthetic constant treatment effect of magnitude $\alpha$ is applied as
\begin{equation*}
    \text{glucose}_{it} = \text{glucose}_{it} - \alpha,
\end{equation*}
if sample $i$ is treated and $t \in I$. We randomize treatment assignment with probability of 0.5 of a sample receiving a treatment after matching samples with the \texttt{nbpmatching} package in R.

\begin{figure}
    \centering
    \includegraphics[scale = 0.65]{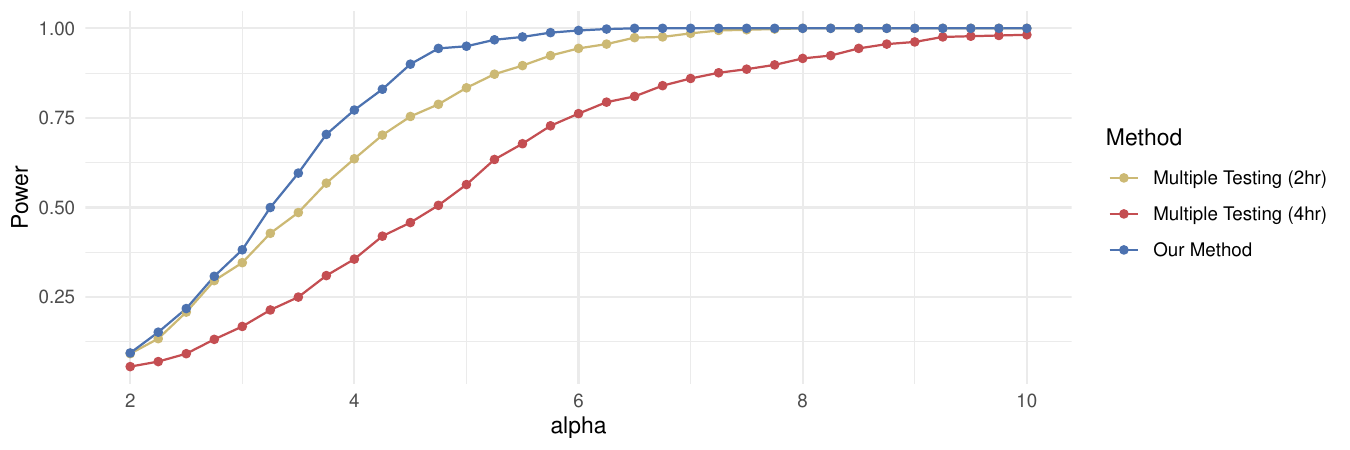}
    \caption{Power plots for the semi-synthetic glucose data comparing our proposed method (blue) to a default method of looking at pre-specified 4-hour windows (red) or 2-hour windows (yellow) and then applying multiple testing.}
    \label{fig:power-plot}
\end{figure}

Suppose the scientist is uncertain about the location and duration of the treatment effect.  One approach they might consider is as follows.  First, they might examine the time-in-range (TIR) in 6 four-hour windows, each roughly corresponding to \emph{early morning}, \emph{late morning}, \emph{afternoon}, \emph{evening}, \emph{night}, and \emph{overnight}. They may then decide they prefer a more refined representation and consider TIR in 12 two-hour windows. Suppose in each scenario the scientist compares the differences in TIR before and after the treatment for each window and conducts multiple testing. As previously discussed in Section \ref{sec:introduction}, if the actual treatment duration is more localized than the window size under consideration, as in the first scenario considering 4-hour windows, it is highly likely that the treatment signal will be lost. The second representation happens to be a fortuitous selection as the actual duration of the simulated treatment effect is two hours; however, as we illustrate in Figure~\ref{fig:power-plot}, our proposed method still outperforms the default multiple testing approach the scientist would employ. 

In contrast to using a fixed length of windows, our proposed approach allows for exploration of more refined windows while achieving better power. In the first stage, we randomly partition the data set into two splits. Within the first split, we aim to locate the treatment effect and consider multiple resolutions for windows. We denote 6 four-hour windows as \emph{level 0}, 12 two-hour windows as \emph{level 1}, and 24 one-hour windows as \emph{level 2}. We perform subset selection with our proposed method where we set the pre-treament covariates $X$ to be TIR of considered windows of the pre-treatment week (Week 1) and $Y$ to be TIR of considered windows of the post-treatment week (Week 2). We perform subset selection for each level and compute the RSS as described in Section \ref{sec:wls} for the selected subset within the level. The level and corresponding selected subset with the smallest RSS is chosen.  Moving to the second split, we estimate the treatment effect utilizing the selected subset of the chosen level. We apply the multi-sample splitting procedure of \cite{meinshausen2009pvalues} to our entire process to reduce the dependence on the random data split; see Algorithm~\ref{algo:multi-split}. 

The entire process, involving the generation of a random two-hour treatment effect and the application of methods introduced above, is repeated 1000 times. The powers of the considered methods from these 1000 iterations are presented in Figure~\ref{fig:power-plot}. We see that our method achieves better power than the default method of using fixed-sized windows and multiple testing. Our method performs better for several reasons: (1) the treatment effect is sparse, and our approach efficiently captures this sparsity by allowing for flexibility in the identification of the treatment effect windows; (2) when the fixed-window size is too large compared to the actual treatment effect, the signal is diluted, leading to reduced power in the baseline method; and (3) even when the fixed-window size is as narrow as the actual sparse treatment effects, the baseline method still suffers from power loss due to the multiple testing corrections applied to each window. Our method, however, can identify the treatment effect window from the first stage and proceed without substantial multiple testing corrections. If multiple windows are selected in the first stage in our method, a multiple testing correction factor is still applied, but it is generally smaller than that required by the baseline method.

\section{Discussion}

This paper addresses the problem of causal inference from experimental data with high-dimensional outcomes.  Our emphasis is on the development of a practical, easily deployed method that can identify the sparse subset of outcome summaries most likely to contain the treatment effect. We first cast the problem as one of maximizing an objective based on the Hotelling's $T^2$ test statistic; to get to computational tractability, we then equate this objective to performing subset selection within a weighted linear regression setting with carefully chosen weights. For this task, we deploy $\ell_1$ regularization. We provide theoretical results showing that our method can recover the appropriate subset of outcome summaries consistently. In simulated and semi-synthetic experiments, we empirically demonstrate that our method outperforms the baseline methods, especially when the signal-to-noise ratio is relatively weak.

A key challenge lies when one wants to employ our method with sample-splitting.  
Sample-splitting based methods may suffer from reduced power as they utilize only half of the data for the treatment effect estimation. Compared to the baseline (individual-ranking-based) subset selection method, our method performs better in low sample-size settings, suggesting the potential for allocating more samples to the second data split for the actual treatment effect estimation. Nonetheless, the overall process necessitates sufficient sample sizes. In settings where the primary challenge is obtaining adequate sample sizes, alternative approaches beyond sample splitting may be required. That said, our procedure is particularly adept if the dimensionality of the outcome summaries far exceeds sample sizes even when sample sizes are sufficient for the second-stage inference (assuming sparsity in the first stage); this is a common scenario in analysis of gene expression data. In clinical settings, which are our main motivation, we often begin with a small pilot or POC study, followed by a much larger clinical trial. In these cases, where the second stage involves a much larger sample size, our method can be effectively applied to select relevant outcome summaries and estimate treatment effects.

\section{Acknowledgments}

The authors would like to thank Dominik Rothenhausler, Kevin Guo, James Yang, Samir Khan, Yu Wang, and Tim Morrison for their helpful feedback. The authors would also like to thank Johannes Ferstad for assistance with the dataset used in the experiments above. This work was supported in part by the National Science Foundation Grant 2205084; the Helmsley Charitable Trust; NIH/NIDDK Grant R18DK122422; AFOSR Grant FA9550-21-1-0397, ONR Grant N00014-22-1-2110; the Stanford Institute for Human-Centered AI (HAI); and the Stanford Maternal and Child Health Research Institute. EBF is a Chan Zuckerberg Biohub - San Francisco Investigator.  YJ is supported by the Stanford Data Science Scholars Program.

\newpage

\bibliography{bibliography}

\newpage
\appendix

\section*{Appendix}

Section~\ref{sec: calculations} contains calculations for Section \ref{sec:wls} and Section~\ref{sec:proofs} contains the proofs for Section \ref{sec:recovery}. 

\section{Calculations for Section \ref{sec:wls}}\label{sec: calculations}

Let us first show the equation \eqref{eq:rss-power}. Note that
\begin{align}
    &\sum_i (T_i - (Y^S_i- \overline{Y^S})^{\intercal}\hat{\beta})^2 \\
    &= \sum T_i - \sum T_i(Y^S_i - \overline{ Y^S})^{\intercal}\left(\sum(Y^S_i - \overline{Y^S})(Y^S_i - \overline{Y^S})^{\intercal} \right)^{-1}\sum T_i(Y^S_i - \overline{Y^S}).
\end{align}
We have that 
\begin{align*}
    \sum T_i(Y^S_i - \overline{Y^S}) &= \sum T_i Y^S_i - \overline{Y^S} \sum T_i \\
    & = \sum T_i Y^S_i(1) - \frac{n_t}{n} \left(\sum T_iY^S_i(1) + \sum (1-T_i)Y^S_i(0)\right) \\
    & = \frac{n_tn_c}{n}\left(\frac{1}{n_t} \sum T_i Y^S_i(1) - \frac{1}{n_c}\sum (1-T_i)Y^S_i(0)\right) .
\end{align*}
Moreover, we get that
\begin{align*}
    &\frac{1}{n}\sum(Y^S_i - \overline{Y^S})(Y^S_i - \overline{Y^S})^{\intercal} = \frac{1}{n}\sum Y^S_i {Y^S_i}^{\intercal}- \overline{Y^S} \overline{Y^S}^{\intercal} \\
    & = \frac{1}{n}\sum T_iY^S_i(1) Y^S_i(1)^{\intercal} + \frac{1}{n}\sum (1-T_i) Y^S_i(0) Y^S_i(0)^{\intercal} - \overline{Y^S} \overline{Y^S}^{\intercal} \\
    & = \frac{n_t}{n} \left(\frac{1}{n_t} \sum T_iY^S_i(1) Y^S_i(1)^{\intercal} - \left(\frac{1}{n_t} \sum T_iY^S_i(1)\right)\left(\frac{1}{n_t} \sum T_iY^S_i(1)\right)^{\intercal}\right) \\
    & + \frac{n_c}{n} \left(\frac{1}{n_c} \sum (1-T_i)Y^S_i(0) Y^S_i(0)^{\intercal} - \left(\frac{1}{n_c} \sum (1-T_i)Y^S_i(0)\right)\left(\frac{1}{n_c} \sum (1-T_i)Y^S_i(0)\right)^{\intercal}\right) \\
    & + \left(\frac{n_t}{n} - \frac{n_t^2}{n^2}\right) \left(\frac{1}{n_t} \sum T_iY^S_i(1)\right)\left(\frac{1}{n_t} \sum T_iY^S_i(1)\right)^{\intercal} \\
    & - \frac{n_tn_c}{n^2} \left(\frac{1}{n_t} \sum T_iY^S_i(1)\right)\left(\frac{1}{n_c} \sum (1-T_i)Y^S_i(0)\right)^{\intercal} \\
    & - \frac{n_tn_c}{n^2}\left(\frac{1}{n_c} \sum (1-T_i)Y^S_i(0)\right)\left(\frac{1}{n_t} \sum T_iY^S_i(1)\right)^{\intercal} \\
    & + \left(\frac{n_c}{n} - \frac{n_c^2}{n^2}\right) \left(\frac{1}{n_c} \sum (1-T_i)Y^S_i(0)\right)\left(\frac{1}{n_c} \sum (1-T_i)Y^S_i(0)\right)^{\intercal} = \frac{n_tn_c}{n^2} \left(\hat{\Sigma}^S + \hat{\tau}_{\text{DiM}}^S{\hat{\tau}_{\text{DiM}}^S}^{\intercal}\right)
\end{align*}
where $\hat{\tau}_{\text{DiM}}^S$ is defined as in \eqref{eq:dim} and $\hat{\Sigma}^S$ is defined as in \eqref{eq:sigmahat_prev}. Combining the results, we have
\begin{align}
    \frac{1}{n}\sum_i (T_i - (Y^S_i- \overline{Y^S})^{\intercal}\hat{\beta})^2 
    &= \frac{n_t}{n} - \frac{n_tn_c}{n^2} \cdot 
    {\hat{\tau}_{\text{DiM}}^S}^{\intercal}   \left(\hat{\Sigma}^S + \hat{\tau}_{\text{DiM}}^S{\hat{\tau}_{\text{DiM}}^S}^{\intercal}\right)^{-1} \hat{\tau}_{\text{DiM}}^S \\
    & = \frac{n_t}{n} - \frac{n_tn_c}{n^2} \cdot \frac{{\hat{\tau}_{\text{DiM}}^S}^{\intercal}(\hat{\Sigma}^S)^{-1}\hat{\tau}_{\text{DiM}}^S}{1 + {\hat{\tau}_{\text{DiM}}^S}^{\intercal}(\hat{{\Sigma}}^S)^{-1}\hat{\tau}_{\text{DiM}}^S},
\end{align}
where we apply Sherman-Morrison formula for the second equality. 

\quad\\
Now let us show the equation \eqref{eq:wls-rss-power}. Note that
\begin{align}
    &\sum_i W_i (T_i - (Y^S_i- \overline{Y^S})^{\intercal}\hat{\beta})^2 \\
    &= \sum W_i T_i - \sum W_iT_i(Y^S_i - \overline{ Y^S})^{\intercal}\left(\sum W_i (Y^S_i - \overline{Y^S})(Y^S_i - \overline{Y^S})^{\intercal} \right)^{-1}\sum W_i T_i(Y^S_i - \overline{Y^S}).
\end{align}
We have that 
\begin{align*}
    \frac{1}{n}\sum W_i T_i(Y^S_i - \overline{Y^S}) = \frac{n_c}{n_t}\left(\frac{1}{n_t} \sum T_i Y^S_i(1) - \frac{1}{n_c}\sum (1-T_i)Y^S_i(0)\right) .
\end{align*}
Moreover, we get that
\begin{align}
    &\frac{1}{n}\sum W_i(Y^S_i - \overline{Y^S})(Y^S_i - \overline{Y^S})^{\intercal} \\
    &= \frac{1}{n}\sum W_i Y^S_i {Y^S_i}^{\intercal}- \left(\frac{1}{n} \sum W_i Y_i^S \right)\overline{Y^S}^{\intercal} - \overline{Y^S}\left(\frac{1}{n} \sum W_i Y_i^S \right)^{\intercal} + \left(\frac{1}{n}\sum W_i\right)\overline{Y^S}\overline{Y^S}^{\intercal} \\
    & = \frac{n}{n_t} \left(\frac{1}{n_t} \sum T_iY^S_i(1) Y^S_i(1)^{\intercal} - \left(\frac{1}{n_t} \sum T_iY^S_i(1)\right)\left(\frac{1}{n_t} \sum T_iY^S_i(1)\right)^{\intercal}\right) \\
     & + \frac{n}{n_c} \left(\frac{1}{n_c} \sum (1-T_i)Y^S_i(0) Y^S_i(0)^{\intercal} - \left(\frac{1}{n_c} \sum (1-T_i)Y^S_i(0)\right)\left(\frac{1}{n_c} \sum (1-T_i)Y^S_i(0)\right)^{\intercal}\right) \\
     & + \left(\frac{n}{n_t} - 2W(1)\frac{n_t^2}{n^2} + \left(\frac{n}{n_t} + \frac{n}{n_c}\right)  \frac{n_t^2}{n^2}\right) \left(\frac{1}{n_t} \sum T_iY^S_i(1)\right)\left(\frac{1}{n_t} \sum T_iY^S_i(1)\right)^{\intercal} \\
     & - \left( W(1) \frac{n_tn_c}{n^2} + W(0) \frac{n_tn_c}{n^2} - \left(\frac{n}{n_t} + \frac{n}{n_c}\right)\frac{n_tn_c}{n^2}\right) \left(\frac{1}{n_t} \sum T_iY^S_i(1)\right)\left(\frac{1}{n_c} \sum (1-T_i)Y^S_i(0)\right)^{\intercal} \\
     & - \left( W(0) \frac{n_tn_c}{n^2} + W(1) \frac{n_tn_c}{n^2} - \left(\frac{n}{n_t} + \frac{n}{n_c}\right)\frac{n_tn_c}{n^2}\right)\left(\frac{1}{n_c} \sum (1-T_i)Y^S_i(0)\right)\left(\frac{1}{n_t} \sum T_iY^S_i(1)\right)^{\intercal} \\
     & + \left(\frac{n}{n_c} - 2W(0)\frac{n_c^2}{n^2} + \left(\frac{n}{n_t} + \frac{n}{n_c}\right)  \frac{n_c^2}{n^2}\right) \left(\frac{1}{n_c} \sum (1-T_i)Y^S_i(0)\right)\left(\frac{1}{n_c} \sum (1-T_i)Y^S_i(0)\right)^{\intercal} \\
     & = \hat{\Sigma}_{\text{DiM}}^S + \left(\frac{n_c}{n_t} + \frac{n_t}{n_c} - 1\right) \hat{\tau}_{\text{DiM}}^S{\hat{\tau}_{\text{DiM}}^S}^{\intercal} \label{eq:WYY}
\end{align}
where $\hat{\tau}_{\text{DiM}}^S$ is defined as in \eqref{eq:dim} and $\hat{\Sigma}_{\text{DiM}}^S$ is defined as in \eqref{eq:sigma-dim}. Combining the results, we have
\begin{align}
    \frac{1}{n}\sum_i W_i (T_i - (Y^S_i- \overline{Y^S})^{\intercal}\hat{\beta})^2 
    &= \frac{n}{n_t} - \frac{n_c^2}{n_t^2} \cdot 
    {\hat{\tau}_{\text{DiM}}^S}^{\intercal}   \left(\hat{\Sigma}_{\text{DiM}}^S + \left(\frac{n_c}{n_t} + \frac{n_t}{n_c} - 1\right) \hat{\tau}_{\text{DiM}}^S{\hat{\tau}_{\text{DiM}}^S}^{\intercal}\right)^{-1} \hat{\tau}_{\text{DiM}}^S \\
    & = \frac{n_t}{n} - \frac{n_c^2}{n_t^2} \cdot \frac{{\hat{\tau}_{\text{DiM}}^S}^{\intercal}(\hat{\Sigma}_{\text{DiM}}^S)^{-1}\hat{\tau}_{\text{DiM}}^S}{1 + \left(\frac{n_c}{n_t} + \frac{n_t}{n_c} - 1\right) \cdot {\hat{\tau}_{\text{DiM}}^S}^{\intercal}(\hat{{\Sigma}}_{\text{DiM}}^S)^{-1}\hat{\tau}_{\text{DiM}}^S},
\end{align}
where we apply Sherman-Morrison formula for the second equality. 

\section{Proofs for Section \ref{sec:recovery}}\label{sec:proofs}

\subsection{RE conditions for sample covariance matrices}\label{sec:re-sample}

Let the sample covariance matrix of $Z$ for the treated group be 
\begin{align*}
    \hat{\Sigma}_Z^1 &= \frac{1}{n_t} \sum_{i:T_i = 1} \left(Z_i - \frac{1}{n_t} \sum_{i:T_i = 1} Z_i(1)\right)\left(Z_i - \frac{1}{n_t}\sum_{i:T_i = 1} Z_i(1)\right)^{\intercal}.
\end{align*}
The sample covariance matrix of $Z$ for the controlled group, $\hat{\Sigma}_Z^0$ is defined similarly.

Set $0 < \theta< 1$. Assume $r\cdot n \geq C\cdot s^* \log p$ for some sufficiently large $C>0$. Suppose Assumption~\ref{ass:sub-gaussianity} and Assumption~\ref{ass:re} hold. Using Theorem 1.6 in \cite{zhou2009restricted}, we have that
$\hat{\Sigma}_Z^1$ satisfies the RE $(s^*, 3, \hat{\Sigma}_Z^1)$ condition with parameter $\kappa_1 := \kappa(s^*, 3, \hat{\Sigma}_Z^1) = (1-\theta)\cdot \kappa(s^*, 3, \Sigma_Z^1)$ and $\hat{\Sigma}_Z^0$ satisfies the RE $(s^*, 3, \hat{\Sigma}_Z^0)$ condition with parameter $\kappa_0 := \kappa(s^*, 3, \hat{\Sigma}_Z^0) = (1-\theta)\cdot \kappa(s^*, 3, \Sigma_Z^0)$,
with probability at least $1-2\exp(-c\theta^2n)$ for some constant $c > 0$.

\subsection{Concentration inequalities for proofs}\label{sec:concentration-inequalities}

In this section, we study concentration inequalities that will be useful for proving Theorem \ref{main-theorem}. First, consider the matrix, 
\begin{equation*}
     \frac{1}{n}\sum_{i=1}^{n} W_i(Z_i -\overline{Z})(Z_i - \overline{Z})^{\intercal} = \frac{n}{n_t}\hat{\Sigma}_Z^1 + \frac{n}{n_c} \hat{\Sigma}_Z^0 + \left(\frac{n_c}{n_t} + \frac{n_t}{n_c} - 1\right)\hat{\tau}_Z {\hat{\tau}_Z }^{\intercal},
\end{equation*}
where the equality comes from calculations in Section \ref{sec: calculations}. Note that 
\begin{align}
    &\Big|\Big| \frac{1}{n}\sum_{i=1}^{n} W_i(Z_i -\overline{Z})(Z_i - \overline{Z})^{\intercal} - \frac{1}{\pi}\Sigma_Z^1 - \frac{1}{1-\pi} \Sigma_Z^0 - \left(\frac{1-\pi}{\pi} + \frac{\pi}{1-\pi} - 1\right)\tau_Z \tau_Z^{\intercal}\Big|\Big|_{\infty} \label{eq:WZZ}\\
    & \leq \Big|\Big| \frac{n}{n_t} \hat{\Sigma}_Z^1  - \frac{1}{\pi}\Sigma_Z^1 \Big|\Big|_{\infty} +\Big|\Big| \frac{n}{n_c} \hat{\Sigma}_Z^0  - \frac{1}{1-\pi}\Sigma_Z^0 \Big|\Big|_{\infty} \\
    & + \Big|\Big| \left(\frac{n_c}{n_t} + \frac{n_t}{n_c} - 1\right)\hat{\tau}_Z {\hat{\tau}_Z }^{\intercal} - \left(\frac{1-\pi}{\pi} + \frac{\pi}{1-\pi} - 1 \right) \tau_Z \tau_Z^{\intercal}\Big|\Big|_{\infty},
\end{align}
where $\tau_Z = E[Z(1) - Z(0)]$. In the following, we obtain the concentration inequality for \eqref{eq:WZZ}.

By Hoeffding's inequality, we have
\begin{equation*}
    \mathbb{P}\left(\Big|\frac{n_t}{n} - \pi \Big| > t_n\right) \leq 2\exp(-2nt_n^2).
\end{equation*}
Then with probability at least $1 - 2\exp(-2nt_n^2)$, we have 
\begin{equation*}
    \Big| \frac{n}{n_t} - \frac{1}{\pi} \Big| = \Big|\frac{n}{n_t}\left(\frac{n_t}{n} - \pi\right) \frac{1}{\pi}\Big| \leq \frac{1}{\pi - \Big|\frac{n_t}{n} - \pi\Big|}\cdot \frac{1}{\pi}\cdot \Big|\frac{n_t}{n} - \pi\Big| \leq \frac{t_n}{\pi^2/2}
\end{equation*}
for sufficiently large $n$ with $t_n \xrightarrow[]{} 0$. Similarly, we can show that 
\begin{equation}\label{eq:weights-bound}
    \Big| \frac{n}{n_t} - \frac{1}{\pi} \Big|, \Big| \frac{n}{n_c} - \frac{1}{1-\pi} \Big|, \Big| \frac{n_c}{n_t} - \frac{1-\pi}{\pi} \Big|, \Big| \frac{n_t}{n_c} - \frac{\pi}{1-\pi} \Big| \leq C\cdot t_n
\end{equation}
for some constant $C> 0$ with probability at least $1 - 2\exp(-2nt_n^2)$.

Back to \eqref{eq:WZZ}, by triangular inequality, we get
\begin{equation*}
    \Big|\Big| \frac{n}{n_t} \hat{\Sigma}_Z^1  - \frac{1}{\pi}\Sigma_Z^1 \Big|\Big|_{\infty} \leq\Big| \frac{n}{n_t} - \frac{1}{\pi} \Big| \cdot  ||\Sigma_Z^1||_{\infty}+ \frac{1}{\pi} ||\hat{\Sigma}_Z^1 - \Sigma_Z^1||_{\infty} + \Big| \frac{n}{n_t} - \frac{1}{\pi} \Big| \cdot  ||\hat{\Sigma}_Z^1 - \Sigma_Z^1||_{\infty}.
\end{equation*}
Similarly, we can bound the other terms in \eqref{eq:WZZ}. This tells us that we now need tail bounds for
\begin{equation*}
    ||\hat{\tau}_Z - \tau||_{\infty}, ||\hat{\Sigma}_Z^0 - \Sigma_Z^0||_{\infty}, ||\hat{\Sigma}_Z^1 - \Sigma_Z^1||_{\infty}, ||\hat{\tau}_Z\hat{\tau}_Z^{\intercal} - \tau\tau^{\intercal}||_{\infty}.
\end{equation*}
Using the Assumption \ref{ass:sub-gaussianity}, for each $j = 1, \dots, d$, we have the following tail bound,
\allowdisplaybreaks
\begin{align}
    \mathbb{P}\left(|\hat{\tau}_{Zj} - \tau_j| \geq t_n\right) & \leq \mathbb{P}\left(| \overline{Z_j(1)} - \mathbb{E}[Z_j(1)] | + | \overline{Z_j(0)} - \mathbb{E}[Z_j(0)] |\geq t_n\right) \\
    & \leq \mathbb{P}\left(| \overline{Z_j(1)} - \mathbb{E}[Z_j(1)] | \geq t_n/2\right) + \mathbb{P}\left(| \overline{Z_j(0)} - \mathbb{E}[Z_j(0)] |\geq t_n/2\right) \\
    & \leq 2 \exp \left( -\frac{n_t t_n^2}{8 (\sigma_Z^1)^2 (\Sigma_Z^1)_{jj}}\right) + 2 \exp \left( -\frac{n_c t_n^2}{8 (\sigma_Z^0)^2 (\Sigma_Z^0)_{jj}}\right) \\
    & \leq 4  \exp \left(- \frac{r n t_n^2}{8((\sigma_Z^1)^2 \max_j (\Sigma_Z^1)_{jj} + (\sigma_Z^0)^2 \max_j (\Sigma_Z^0)_{jj})}\right) \label{eq:sub-tail-bound} .
\end{align}
By union bound, we have that 
\begin{align}
    \mathbb{P} \left(||\hat{\tau}_Z  - \tau ||_{\infty} \geq t_n \right) 
    &\leq 4  \exp \left(- \frac{r n t_n^2}{8((\sigma_Z^1)^2 \max_j (\Sigma_Z^1)_{jj} + (\sigma_Z^0)^2 \max_j (\Sigma_Z^0)_{jj})}+ \log p \right). \label{eq: bound1}
\end{align}
Using the Assumption \ref{ass:sub-gaussianity} and Lemma 1 in \cite{ravikumar2011cov}, we again have the tail bound 
\begin{equation*}
    \mathbb{P}\left(|(\hat{\Sigma}_Z^1)_{ij} - (\Sigma_Z^1)_{ij}| \geq t_n \right) \leq 4 \exp\left(- \frac{r n t_n^2}{128(1 + 4(\sigma_Z^1)^2)^2\max_j((\Sigma_Z^1)_{jj})^2}\right).
\end{equation*}
By union bound, we have that 
\begin{align}
    \mathbb{P}\left(||\hat{\Sigma}_Z^1 - \Sigma_Z^1||_{\infty} \geq t_n \right) \leq 4 \exp\left(- \frac{r n t_n^2}{128(1 + 4(\sigma_Z^1)^2)^2\max_j((\Sigma_Z^1)_{jj})^2} + 2 \log p \right). \label{eq: bound2}
\end{align}
Similarly, we have that 
\begin{align}
    \mathbb{P}\left(||\hat{\Sigma}_Z^0 - \Sigma_Z^0||_{\infty} \geq t_n \right) \leq 4 \exp\left(- \frac{r n t_n^2}{128(1 + 4(\sigma_Z^0)^2)^2\max_j((\Sigma_Z^0)_{jj})^2} + 2 \log p \right). \label{eq: bound3}
\end{align}
Moreover, we get
\begin{align*}
\mathbb{P}\left(|\hat{\tau}_{Zi} \hat{\tau}_{Zj}  - \tau_i \tau_j | \geq t_n\right) &\leq \mathbb{P}\left(|\hat{\tau}_{Zi} - \tau _i| \cdot |\hat{\tau}_{Zj} - \tau _j| + |\tau_i |\cdot|\hat{\tau}_{Zj} - \tau _j| + |\tau_j | \cdot |\hat{\tau}_{Zi} - \tau _i| \geq t_n \right) \\
&\leq \mathbb{P}(|\hat{\tau}_{Zi} - \tau _i| \cdot |\hat{\tau}_{Zj} - \tau _j| \geq t_n/3) \\ & \quad + \mathbb{P}(|\tau_i |\cdot|\hat{\tau}_{Zj} - \tau _j| \geq t_n/3) + \mathbb{P}(|\tau_j | \cdot |\hat{\tau}_{Zi} - \tau _i| \geq t_n/3) \\
& \leq  8 \exp \left(- \frac{r n t_n}{6( (\sigma_Z^1)^2 \max_j (\Sigma_Z^1)_{jj} + (\sigma_Z^0)^2 \max_j (\Sigma_Z^0)_{jj})}\right) \\ 
& + 8 \exp \left(- \frac{r n t_n^2}{18 \max_j |\tau_j |^2( (\sigma_Z^1)^2 \max_j (\Sigma_Z^1)_{jj} + (\sigma_Z^0)^2 \max_j (\Sigma_Z^0)_{jj})}\right) I\left(\max_j |\tau_j | > 0\right).
\end{align*}
Similarly, we apply the union bound to get the following tail bound 
\begin{align}
    &\mathbb{P}\left( ||\hat{\tau}_Z {\hat{\tau} }_Z^{\intercal} - {\tau} {{\tau} }^{\intercal}||_{\infty} \geq t_n \right) \\ & \leq  8 \exp \left(- \frac{r n t_n}{6( (\sigma_Z^1)^2 \max_j (\Sigma_Z^1)_{jj} + (\sigma_Z^0)^2 \max_j (\Sigma_Z^0)_{jj})} + 2\log p\right) \\ 
& + 8 \exp \left(- \frac{r n t_n^2}{18 \max_j |\tau_j |^2( (\sigma_Z^1)^2 \max_j (\Sigma_Z^1)_{jj} + (\sigma_Z^0)^2 \max_j (\Sigma_Z^0)_{jj})} + 2\log p\right) I\left(\max_j |\tau_j | > 0\right). \label{eq: bound4}
\end{align}

Let $t_n \geq A\sqrt{\log p /n}$ where $A >1$ is sufficiently large enough to ensure that exponential factors in \eqref{eq: bound1}, \eqref{eq: bound2}, \eqref{eq: bound3}, and \eqref{eq: bound4} are negative. Combining results in \eqref{eq:weights-bound}, \eqref{eq: bound1}, \eqref{eq: bound2}, \eqref{eq: bound3}, and \eqref{eq: bound4}, we have
\begin{align*}
    \eqref{eq:WZZ} \text{ } \leq C \cdot t_n
\end{align*}
with probability at least $1 - c_1 \exp(-c_2 n t_n^2)$ for some constants $C, c_1, c_2 > 0$. 

Similarly, we can get
\begin{align}
    & \Big|\Big| \frac{1}{n}\sum_{i=1}^{n} W_i(Z_i -\overline{Z})(X_i - \overline{X})^{\intercal} - \frac{1}{\pi}\text{Cov}(Z(1), X) - \frac{1}{1-\pi}\text{Cov}(Z(0), X)\Big|\Big|_{\infty}  \\& =  \Big|\Big| \frac{1}{n}\sum_{i=1}^{n} W_i(Z_i -\overline{Z})(X_i - \overline{X})^{\intercal} \Big|\Big|_{\infty} \leq C \cdot t_n \label{eq:WZX}\\
    & \Big|\Big| \frac{1}{n}\sum_{i=1}^{n} W_i(X_i -\overline{X})(X_i - \overline{X})^{\intercal} - \frac{1}{\pi(1-\pi)}\Sigma_X \Big|\Big|_{\infty}\leq C \cdot t_n \label{eq:WXX}
\end{align}
with probability at least $1 - c_1 \exp(-c_2 n t_n^2)$ for some constants $C, c_1, c_2 > 0$. 

Moreover, since the pre-treatment covariates dimension $m$ is fixed, using the matrix Bernstein inequality \citep{Wainwright_2019}, we get
\begin{align} 
    &\Big|\Big| \frac{1}{n}\sum_{i=1}^{n} W_i(X_i -\overline{X})(X_i - \overline{X})^{\intercal} - \frac{1}{\pi(1-\pi)}\Sigma_X \Big|\Big|_2 \leq C \cdot t_n\label{eq:regbounds-ineq-1} \\
    & \Big|\Big| \frac{1}{n}\sum_{i=1}^{n} W_iT_i(X_i -\overline{X}) \Big|\Big|_2\leq C \cdot t_n \label{eq:regbounds-ineq-2}\\
    & \Big|\Big| \frac{1}{n}\sum_{i=1}^{n} W_i(X_i -\overline{X})(Y_{ij} - \overline{Y}_j)^{\intercal} - \frac{1}{\pi}\text{Cov}(X, Y_j(1)) - \frac{1}{1-\pi}\text{Cov}(X, Y_j(0)) \Big|\Big|_2\leq C \cdot t_n \label{eq:regbounds-ineq-3}
\end{align}
for $j = 1, \dots, p$, 
with probability at least $1 - c_1 \exp(-c_2 n t_n^2)$ for some constants $C, c_1, c_2 > 0$.

\subsection{Proof of Theorem \ref{main-theorem}}

Note that 
\begin{equation*}
    \hat{\alpha} = \left[\frac{1}{n}\sum_{i=1}^n W_i (X_i - \Bar{X})(X_i - \Bar{X})^{\intercal}\right]^{-1}\left(\frac{1}{n}\sum_{i=1}^n W_i(X_i - \Bar{X})T_i - \frac{1}{n}\sum_{i=1}^n W_i(X_i - \Bar{X})(Y_i - \Bar{Y})^{\intercal}\beta\right).
\end{equation*}
Let 
\begin{align*}
    \hat{b} & =  \left[\frac{1}{n}\sum_{i=1}^n W_i (X_i - \Bar{X})(X_i - \Bar{X})^{\intercal}\right]^{-1}\frac{1}{n}\sum_{i=1}^n W_i(X_i - \Bar{X})T_i\\
    \hat{A} & =  \left[\frac{1}{n}\sum_{i=1}^n W_i (X_i - \Bar{X})(X_i - \Bar{X})^{\intercal}\right]^{-1}\frac{1}{n}\sum_{i=1}^n W_i(X_i - \Bar{X})(Y_i - \Bar{Y})^{\intercal}.
\end{align*}
The following lemma gives the concentration inequalities for $\hat{b}$ and $\hat{A}$. The proof of the Lemma \ref{lemma:regbound} can be found in Section \ref{sec:lemma-1}.
\begin{lemma}\label{lemma:regbound}
    There exist positive constants $b_1, b_2, a_1, a_2$ such that
    \begin{align}
        &\mathbb{P}\left(||\hat{b}||_1 \geq t_n \right) \leq b_1 \exp\left(- b_2 nt_n^2 \right)\\ &\mathbb{P}\left(\max_{j=1}^p||\hat{A}_j - A_j||_1 \geq t_n \right) \leq a_{1} \exp\left(- a_{2} nt_n^2 + \log p\right).
    \end{align}
   where $\hat{A}_j$ is the $j$-th column of $\hat{A}$ and $A_j$ is defined as 
    \begin{equation*}
        A_j := (1-\pi)\text{Var}(X)^{-1}\text{Cov}(X, Y_j(1)) + \pi\text{Var}(X)^{-1}\text{Cov}(X, Y_j(0)).
    \end{equation*}
\end{lemma}

Note that $\hat{\beta}$ can be written as 
\begin{align}
    \hat{\beta} &= \arg\min_{\beta }\frac{1}{n}\sum_i W_i(T_i - (X_i - \Bar{X})^{\intercal} \hat{b} - \left[(Y_i- \overline{Y})- \hat{A}^{\intercal}(X_i - \Bar{X})\right]^{\intercal}{\beta})^2  + 2\lambda_n ||\beta||_1 \\
    &= \arg\min_{\beta }\frac{1}{n}\sum_i W_i(T_i - (X_i - \Bar{X})^{\intercal} \hat{b} - \left[(Z_i- \overline{Z})- (\hat{A}- A)^{\intercal}(X_i - \Bar{X})\right]^{\intercal}{\beta})^2  + 2\lambda_n ||\beta||_1. \label{eq:new_betahat}
\end{align}
From the sub-differential property of the minimizer in \eqref{eq:new_betahat}, we have
\begin{align}
    \Big|\Big| \frac{1}{n}\sum_{i=1}^{n} W_i&\left(Z_i- \overline{Z}- (\hat{A}- A)^{\intercal}(X_i - \Bar{X})\right)\\
    &\left(T_i - (X_i - \Bar{X})^{\intercal} \hat{b} - \left[Z_i- \overline{Z}- (\hat{A}- A)^{\intercal}(X_i - \Bar{X})\right]^{\intercal}\hat{\beta}\right)\Big|\Big|_{\infty} \leq \lambda_n.\label{eq:subdiff_constraint_2}
\end{align}
Set $\bm{\delta} = \hat{\beta} - \beta^*$ and $J_0 = J(\beta^*)$. By Lemma 1 in \cite{negahban2012Lasso}, if 
 \begin{align}
     \Big|\Big| \frac{1}{n}\sum_{i=1}^{n} W_i&\left(Z_i- \overline{Z}- (\hat{A}- A)^{\intercal}(X_i - \Bar{X})\right)\\
     &\left(T_i - (X_i - \Bar{X})^{\intercal} \hat{b} - \left[Z_i- \overline{Z}- (\hat{A}- A)^{\intercal}(X_i - \Bar{X})\right]^{\intercal}{\beta}^*\right)\Big|\Big|_{\infty} \leq \lambda_n, \label{eq:lambda_cond_2}
\end{align}
then $\bm{\delta}$ satisfies
\begin{equation}\label{eq:cone_set_2}
    ||\bm{\delta}_{J_0^c}||_1 \leq 3 ||\bm{\delta}_{J_0}||_1.
\end{equation}

We first state the lemma saying that \eqref{eq:lambda_cond_2} holds with high probability. The proof can be found in Section \ref{sec:lemma-2}.
\begin{lemma}\label{lemma:supp-lemma}
Suppose the assumptions of Theorem \ref{main-theorem} hold. Let $\lambda_n \geq A \sqrt{\log p /n}$ 
for some sufficiently large $A > 0$. Then with probability at least $1 - c_1 \exp(-c_2 n \lambda_n^2)$ for some positive constants $c_1, c_2$, we have the inequality \eqref{eq:lambda_cond_2}.
\end{lemma}

With the choice of $\lambda_n$ in Lemma \ref{lemma:supp-lemma},
from \eqref{eq:subdiff_constraint_2} and \eqref{eq:lambda_cond_2}, we have that
\begin{align*}
     \Big|\Big| \frac{1}{n}\sum_{i=1}^{n} W_i\left(Z_i- \overline{Z}- (\hat{A}- A)^{\intercal}(X_i - \Bar{X})\right)\left(Z_i- \overline{Z}- (\hat{A}- A)^{\intercal}(X_i - \Bar{X})\right)^{\intercal} \bm{\delta}\Big|\Big|_{\infty} \leq 2 \lambda_n,
\end{align*}
with probability at least $1 - c_1 \exp(-c_2 n \lambda_n^2)$ for some positive constants $c_1, c_2 >0$. Then, we get that
\begin{align*} 
     \Big|\Big| \frac{1}{n}\sum_{i=1}^{n} W_i(Z_i- \overline{Z})(Z_i- \overline{Z})^{\intercal}\bm{\delta}\Big|\Big|_{\infty} &\leq
     2 \lambda_n + 2 \Big|\Big| \frac{1}{n}\sum_{i=1}^{n} W_i(Z_i- \overline{Z})(X_i - \overline{X})^{\intercal}(\hat{A} - A)\Big|\Big|_{\infty}||\bm{\delta}||_1 \\
     &+ \Big|\Big| (\hat{A} - A)^{\intercal} \frac{1}{n}\sum_{i=1}^{n} W_i(X_i- \overline{X})(X_i- \overline{X})^{\intercal}(\hat{A} - A)\Big|\Big|_{\infty} ||\bm{\delta}||_1 \\
     & \leq 2\lambda_n + 3\lambda_n^2 ||\bm{\delta}||_1,
\end{align*}
where the last inequality comes from \eqref{eq:WZX}, \eqref{eq:WXX}, and Lemma \ref{lemma:regbound}. Therefore, 
\begin{align*}
    \bm{\delta}^{\intercal}\left( \frac{1}{n}\sum_{i}W_i(Z_i - \overline{Z})(Z_i - \overline{Z})^{\intercal}\right) \bm{\delta} &\leq 
     \Big|\Big| \frac{1}{n}\sum_{i=1}^{n} W_i(Z_i- \overline{Z})(Z_i- \overline{Z})^{\intercal}\bm{\delta}\Big|\Big|_{\infty} ||\bm{\delta}||_1 \\
     &\leq 2\lambda_n||\bm{\delta}||_1 + 3\lambda_n^2 ||\bm{\delta}||_1^2 \\
     & \leq 8\sqrt{s^*}\lambda_n|| \bm{\delta}_{J_0}||_2 + 48 s^* \lambda_n^2|| \bm{\delta}_{J_0}||_2^2.
\end{align*}
From the RE assumption in Section~\ref{sec:re-sample}, we have
\begin{align}
    \bm{\delta}^{\intercal}\left( \frac{1}{n}\sum_{i}W_i(Z_i - \overline{Z})(Z_i - \overline{Z})^{\intercal}\right) \bm{\delta}
    & \geq \bm{\delta}^{\intercal}\left( \frac{n}{n_t} \hat{\Sigma}_Z^1 + \frac{n}{n_c} \hat{\Sigma}_Z^0 \right) \bm{\delta} \\
    & \geq (\kappa_1+ \kappa_0) ||\bm{\delta}_{J_0}||_2^2/(1-r).
\end{align}
Combining these results, we have
\begin{equation*}
     (\kappa_1+ \kappa_0)||\bm{\delta}_{J_0}||_2 \leq  8\sqrt{s^*}\lambda_n + 48 s^* \lambda_n^2|| \bm{\delta}_{J_0}||_2.
\end{equation*}
Given that $n \gg s^*\log p$, we have
\begin{equation*}
     \frac{(\kappa_1+ \kappa_0)}{2}||\bm{\delta}_{J_0}||_2 \leq  8\sqrt{s^*}\lambda_n.
\end{equation*}
Then, we have
\begin{equation*}
    ||\bm{\delta}||_1 = ||\bm{\delta}_{J_0}||_1 + ||\bm{\delta}_{J_0^c}||_1 \leq 4 ||\bm{\delta}_{J_0}||_1 \leq 4\sqrt{s^*}||\bm{\delta}_{J_0}||_2 \leq \frac{64\lambda_n s^*}{\kappa_1 + \kappa_0} .
\end{equation*}
Then the recovery result follows from the fact that the bound on the $\ell_1$-distance gives the identical bound on the $\ell_{\infty}$-distance between $\hat{\beta}$ and $\beta^*$. 
This completes the proof.

\subsection{Proofs of Lemma \ref{lemma:supp-lemma}}\label{sec:lemma-2}
In this Section, we prove Lemma \ref{lemma:supp-lemma}. Note that 
\allowdisplaybreaks
\begin{align}
     &\text{the } \text{left hand side of \eqref{eq:lambda_cond_2}}\\
    & \leq \Big|\Big| \frac{1}{n}\sum_{i=1}^{n} W_i(Z_i -\overline{Z})(T_i - (Z_i - \overline{Z})^{\intercal} {\beta}^*)\Big|\Big|_{\infty} \\
    & + \Big|\Big| \frac{1}{n}\sum_{i=1}^{n} W_i(Z_i -\overline{Z})(X_i - \overline{X})^{\intercal}\hat{b}\Big|\Big|_{\infty} +  \Big|\Big|(\hat{A}- A)^{\intercal} \frac{1}{n}\sum_{i=1}^{n} W_i(X_i -\overline{X})T_i \Big|\Big|_{\infty} \\
    &+  \Big|\Big|(\hat{A}- A)^{\intercal} \frac{1}{n}\sum_{i=1}^{n} W_i(X_i -\overline{X})(X_i - \Bar{X})^{\intercal} \hat{b} \Big|\Big|_{\infty} \\
    & + 2\cdot \Big|\Big| \frac{1}{n}\sum_{i=1}^{n} W_i(Z_i -\overline{Z})(X_i - \overline{X})^{\intercal}(\hat{A}- A)\Big|\Big|_{\infty}||\beta^*||_1\\
    & +   \Big|\Big|(\hat{A}- A)^{\intercal} \frac{1}{n}\sum_{i=1}^{n} W_i(X_i -\overline{X})(X_i - \Bar{X})^{\intercal}(\hat{A}- A)\Big|\Big|_{\infty} ||\beta^*||_1 \\
    &\leq \Big|\Big| \frac{1}{n}\sum_{i=1}^{n} W_i(Z_i -\overline{Z})(T_i - (Z_i - \overline{Z})^{\intercal} {\beta}^*)\Big|\Big|_{\infty} \label{eq:part-1} \\
    &+ ||\hat{b}||_1 \cdot \Big|\Big| \frac{1}{n}\sum_{i=1}^{n} W_i(Z_i -\overline{Z})(X_i - \overline{X})^{\intercal}\Big|\Big|_{\infty} \label{eq:part-2}\\
    &+  \left(\max_{j=1}^d ||\hat{A}_j - A_j||_1\right) \cdot \left( \Big|\Big|\frac{1}{n}\sum_{i=1}^{n} W_i(X_i -\overline{X})T_i \Big|\Big|_{\infty} +  2 \Big|\Big| \frac{1}{n}\sum_{i=1}^{n} W_i(Z_i -\overline{Z})(X_i - \overline{X})^{\intercal}\Big|\Big|_{\infty}||\beta^*||_1\right) \label{eq:part-3}\\
    & + ||\hat{b}||_1 \cdot \left(\max_{j=1}^d ||\hat{A}_j - A_j||_1\right) \cdot \Big|\Big| \frac{1}{n}\sum_{i=1}^{n} W_i(X_i -\overline{X})(X_i - \Bar{X})^{\intercal} \Big|\Big|_{\infty}  \label{eq:part-4}\\
    & +  \left(\max_{j=1}^d ||\hat{A}_j - A_j||_1\right)^2 \cdot  \Big|\Big| \frac{1}{n}\sum_{i=1}^{n} W_i(X_i -\overline{X})(X_i - \Bar{X})^{\intercal}\Big|\Big|_{\infty} ||\beta^*||_1.\label{eq:part-5}
\end{align}

Let's start with \eqref{eq:part-1}. Note that from the calculations in Section \ref{sec: calculations}, we get 
\begin{align} 
    & \Big|\Big| \frac{1}{n}\sum_{i=1}^{n} W_i(Z_i -\overline{Z})(T_i - (Z_i - \overline{Z})^{\intercal} {\beta}^*)\Big|\Big|_{\infty} \\
    & = \Big|\Big| \frac{n_c}{n_t} \hat{\tau}_Z  - \left(\frac{n}{n_t}\hat{\Sigma}_Z^1 + \frac{n}{n_c} \hat{\Sigma}_Z^0 + \left(\frac{n_c}{n_t} + \frac{n_t}{n_c} - 1\right)\hat{\tau}_Z {\hat{\tau}_Z }^{\intercal}\right)\beta^* \Big|\Big|_{\infty}\\
    & \leq \Big|\Big| \frac{n_c}{n_t} \hat{\tau}_Z  - \frac{1-\pi}{\pi}\tau \Big|\Big|_{\infty} +\Big|\Big| \frac{n}{n_t} \hat{\Sigma}_Z^1  - \frac{1}{\pi}\Sigma_Z^1 \Big|\Big|_{\infty} ||\beta^*||_1+\Big|\Big| \frac{n}{n_c} \hat{\Sigma}_Z^0  - \frac{1}{1-\pi}\Sigma_Z^0 \Big|\Big|_{\infty}||\beta^*||_1\\
    & \quad + \Big|\Big| \left(\frac{n_c}{n_t} + \frac{n_t}{n_c} - 1\right)\cdot \hat{\tau}_Z {\hat{\tau}_Z }^{\intercal} - \left(\frac{1-\pi}{\pi} + \frac{\pi}{1-\pi} - 1 \right) {\tau} {{\tau} }^{\intercal}\Big|\Big|_{\infty} ||\beta^*||_1.
 \end{align}
 
 From Section~\ref{sec:concentration-inequalities}, with $t_n \geq A \sqrt{\log p / n}$ for sufficiently large $A > 1$, 
 \begin{equation*}
     \eqref{eq:part-1} \leq M \cdot t_n 
 \end{equation*}
 with probability at least $1 - c_1 \exp(-c_2 n t_n^2)$ for some constants $M, c_1, c_2 > 0$.

Similarly, using results from Section~\ref{sec:concentration-inequalities} and Lemma~\ref{lemma:regbound}, we have that 
\begin{equation*}
    \eqref{eq:part-2},\eqref{eq:part-3}, \eqref{eq:part-4}, \eqref{eq:part-5} \leq 2 t_n^2 (1+||\beta^*||_1)(M +t_n) 
\end{equation*}
with probability at least $1 - c_1 \exp(-c_2 n t_n^2)$ for some constants $M, c_1, c_2 > 0$. 

Eventually, combining all the results, we have
\begin{align}
     \text{the } \text{left hand side of \eqref{eq:lambda_cond_2}} \leq \lambda_n
\end{align}
for $\lambda_n \geq A \sqrt{\log p/n}$ for some sufficiently large $A > 1$, with probability at least  $1 - c_1 \exp(-c_2 n \lambda_n^2)$ for some constants $c_1, c_2 > 0$. This completes the proof. 

\subsection{Proof of Lemma \ref{lemma:regbound}}\label{sec:lemma-1}
Define 
\begin{align*}
    \hat{\Sigma}^W_X &= \pi(1-\pi)\cdot  \frac{1}{n}\sum_{i=1}^n W_i (X_i - \Bar{X})(X_i - \Bar{X})^{\intercal} \\
     \hat{\Sigma}^W_{XY_j} &= \pi(1-\pi)\cdot \frac{1}{n}\sum_{i=1}^n W_i (X_i - \Bar{X})(Y_{ij} - \Bar{Y}_j)^{\intercal}\\
     \Sigma_{XY_j} &= (1-\pi) \text{Cov}(X, Y_j(1)) + \pi \text{Cov}(X, Y_j(0)).
\end{align*}
With probability at least $1 - c_1 \exp(-c_2 n t_n^2)$,  using \eqref{eq:regbounds-ineq-1}, we have
\begin{align*}
    ||{\hat{\Sigma}^W_X}^{-1} - \Sigma_X^{-1}||_{2} &= ||{\Sigma}_X^{-1}(\hat{\Sigma}^W_X - \Sigma_X){\hat{\Sigma}^W_X}^{-1}||_{2} \\
    & \leq ||{\Sigma}_X^{-1}||_{2} ||\hat{\Sigma}^W_X - \Sigma_X||_{2} ||{\hat{\Sigma}^W_X}^{-1}||_{2} \\
    & \leq ||{\Sigma}_X||^{-1}_{2}\cdot t_n \cdot \frac{1}{||{\Sigma}_X||_{2}- t_n } \leq \frac{t_n}{||{\Sigma}_X||^2_{2}/2}
\end{align*}
for sufficiently large $n$ with $t_n \xrightarrow[]{}0$.
Therefore, using \eqref{eq:regbounds-ineq-2}, we have
\begin{equation*}
    ||\hat{b}||_2 \leq ||{\hat{\Sigma}^W_X}^{-1} ||_2\Big|\Big|\pi(1-\pi)\cdot\frac{1}{n}\sum_{i=1}^n W_i T_i(X_i - \Bar{X})\Big|\Big|_2 \leq \frac{t_n}{||\Sigma_X||_{2}/2}
\end{equation*}
and using \eqref{eq:regbounds-ineq-3}, we get
\begin{align*}
    ||\hat{A}_j - A_j||_2 &\leq ||{\hat{\Sigma}^W_X}^{-1} - \Sigma_X^{-1}||_2 ||\Sigma_{XY_j}||_2 + ||\Sigma_X^{-1}||_2 ||\hat{\Sigma}^W_{XY_j} - {\Sigma}_{XY_j}||_2 \\
    & + ||{\hat{\Sigma}^W_X}^{-1} - \Sigma_X^{-1}||_2 ||\hat{\Sigma}^W_{XY_j} - {\Sigma}_{XY_j}||_2 \\
    &  \leq C_1 t_n + C_2 t_n^2,
\end{align*}
for some constants $C_1, C_2 > 0$.
Note that $||x||_1 \leq \sqrt{m}||x||_2$ for $x \in \mathbb{R}^{m}$. This completes the proof.

\section{Multi Sample-Splitting}\label{sec:mss}

\paragraph{Extension to Multi-Sample Splitting.} Methods based on sample splitting are known to be sensitive to the arbitrary choice of a one-time random split of the data \citep{meinshausen2009pvalues}. Therefore, it is recommended to employ multiple splits and aggregate p-values from each split to improve robustness of results. In Algorithm~\ref{algo:multi-split}, we present an algorithm where we apply the multi-sample splitting results from \cite{meinshausen2009pvalues} to our proposed method.  Note that one consequence of using multiple splits is that while we obtain an aggregated p-value for testing the null hypothesis, we no longer obtain an estimated subset localizing the treatment effect (since each split produces one such subset).  Aggregating the subsets remains an interesting direction for future work.

\begin{algorithm}
\DontPrintSemicolon
  \KwInput{$(T_i, X_i, Y_i)_{i=1}^{n}$, $\gamma$}
  \KwOutput{valid p-values under $H_0:\tau = 0$}
  \For{b = 1, \dots, B}{
    Perform Algorithm~\ref{algo:single-split}.
    We have
    $\hat{S}_{\tau}^{(b)}$ from the first split and the treatment effect estimate $\hat{\tau}^{(b)}$ restricted to the set $\hat{S}_{\tau}^{(b)}$ and its variance estimate $\hat{\Sigma}^{(b)}$ from the second split. \\
    Calculate p-value:
    \begin{enumerate}
        \item[(a)] (P-values with multiple testing) The p-value for each $Y_j$ is
        \begin{align*}
             p_j^{(b)} &= \left(1-\Phi\left( |\hat{\tau}^{(b)}_j|/\sqrt{n\hat{\Sigma}^{(b)}_{jj}}\right)\right) \cdot |\hat{S}_{\tau}^{(b)}|\quad \text{for } j \in \hat{S}_{\tau}^{(b)}\\
              p_j^{(b)} &= 1 \quad \text{for } j \notin \hat{S}_{\tau}^{(b)},
        \end{align*}
        where $\Phi$ is the cdf of the standard Gaussian distribution. 
        \item[(b)] (P-value with Hotelling $T^2$ statistic) The p-value is
        \begin{align*}
            p_1^{(b)} = P\left(\chi^2\left(|\hat{S}_{\tau}^{(b)}|\right) \geq n_2 \hat{\tau}^{(b)\intercal} (\hat{\Sigma}^{(b)})^{-1}\hat{\tau}^{(b)}\right),
        \end{align*}
        where $\chi^2(df)$ follows the chi-squared distribution with the degree of freedom $df$.
    \end{enumerate}
  }
  Aggregate p-values as
  \begin{equation*}
      P_j(\gamma) = \min\left[1, q_{\gamma}\left(\{p_j^{(b)}/\gamma; b = 1, \dots, B\}\right)\right],
  \end{equation*}
  where $q_{\gamma}(\cdot)$ is the (empirical) $\gamma$-quantile function. 
\caption{Multi-Split Algorithm}\label{algo:multi-split}
\end{algorithm}

\section{Additional Simulations}\label{sec:add_sims}

\subsection{Without Pre-treament Covarites}

We consider scenarios involving independent outcomes and a constant treatment effect without pre-treatment covariates. We generate the data set as follows. 
\begin{align}
    Y_{ij} = \alpha \cdot T_i  \cdot I(j \leq s^*) + \epsilon_{ij},\\
     T_i \stackrel{i.i.d.}{\sim} \text{Bernoulli}(p),\\
    \epsilon_{ij} \stackrel{i.i.d.}{\sim}  N(0, 1).
\end{align}
for $i = 1, \dots, n$ and $j = 1, \dots, d$. Note that the outcome dimension is $d$, and there are non-zero treatment effects on $S^* = \{Y_1, \dots, Y_{s^*}\}$, where the sparsity index is $s^*$.

\begin{figure}[ht]
    \centering
    \includegraphics[scale = 0.7]{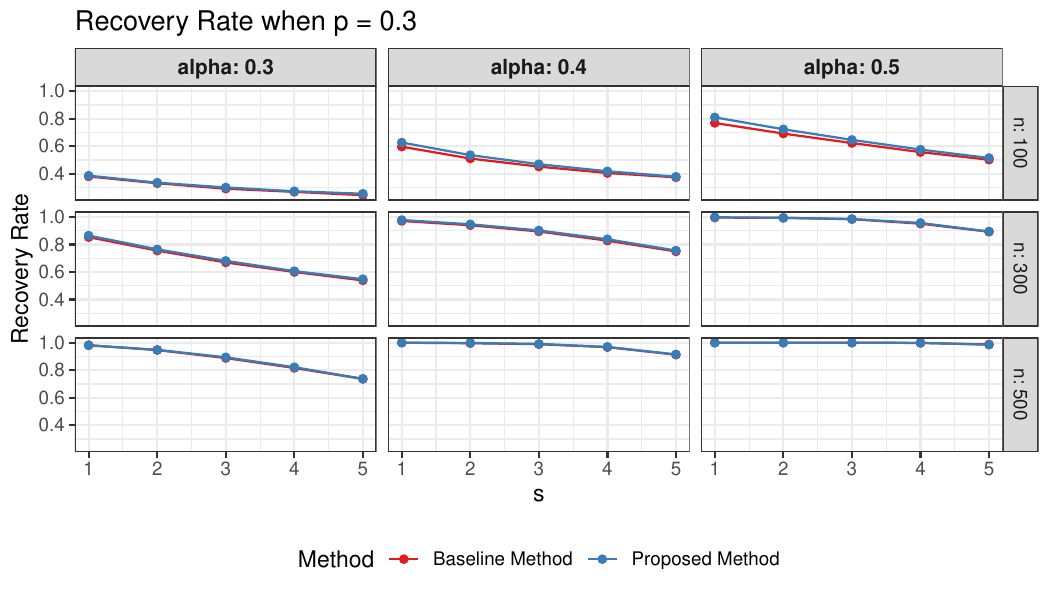}
    \includegraphics[scale = 0.7]{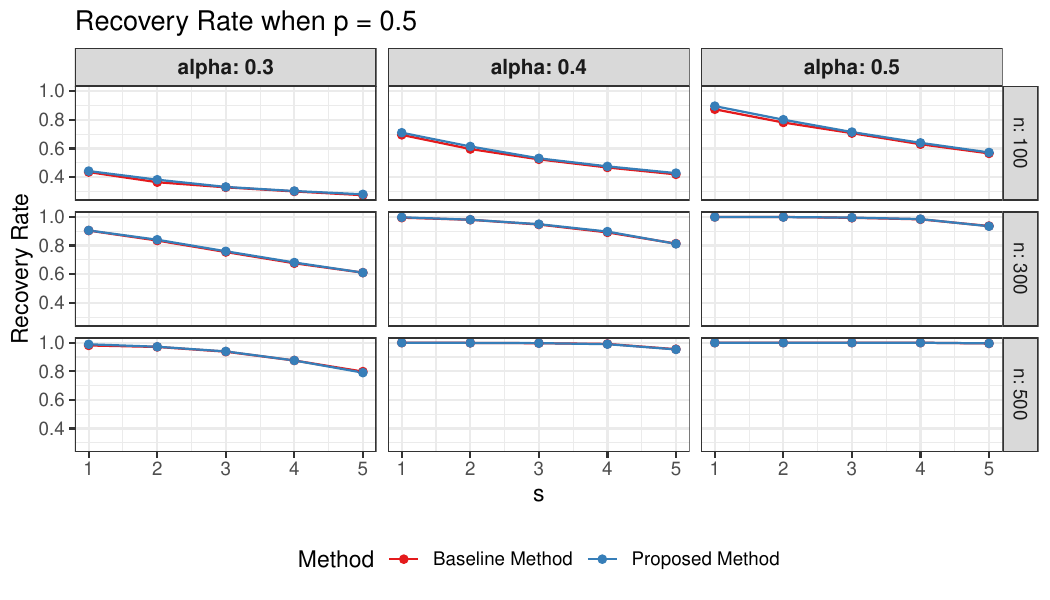}
    \caption{Recovery rates averaged over 1000 replicates for the baseline approach and the proposed approach (without pre-treatment covariates adjustment) for each $n = 100, 300, 500$, $p = 0.3, 0.5$, and $\alpha = 0.3, 0.4, 0.5$ are presented.}
    \label{fig:recovery-rate-no-x}
\end{figure}

The recovery rates of true $S^*$ for the baseline approach and the proposed approach for different sample sizes ($n$), probabilities of being treated ($p$), and magnitudes of treatment effects ($\alpha$) are given in Figure \ref{fig:recovery-rate-no-x}. We set the outcome dimension $d = 100$ and the sparsity index $s^* = 5$. Both the baseline approach and the proposed approach show similar performance.

\subsection{With Partially-Observable Pre-treatment Covariates}

\begin{figure}
    \centering
    \includegraphics[scale = 0.8]{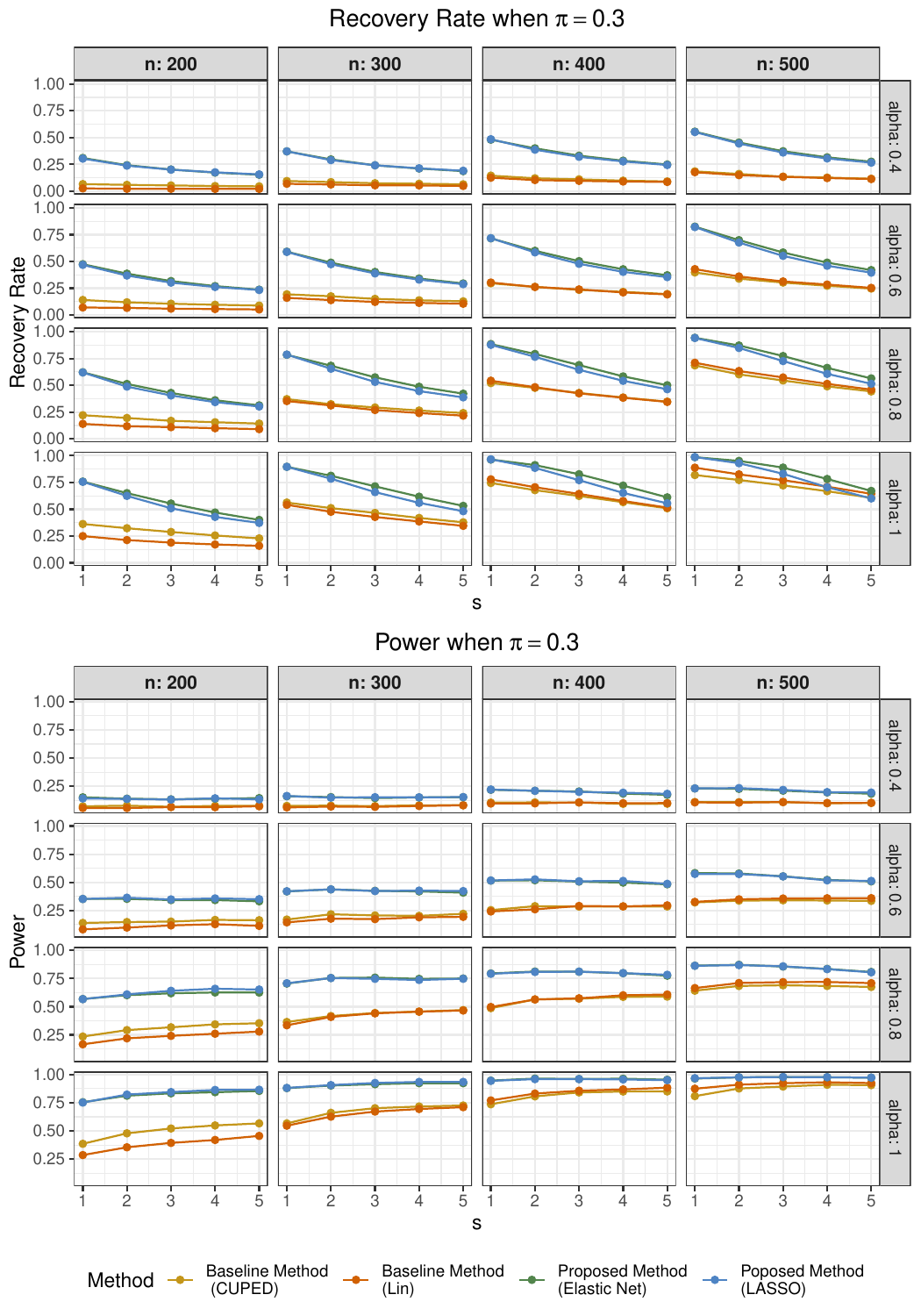}
    \caption{Recovery rates and powers averaged over 1000 replicates for the baseline and proposed approaches with pre-treatment covariates adjustment when pre-treament covariates are partially observable for $n = 200, 300, 400, 500$ and $\alpha = 0.4, 0.6, 0.8, 1$ for $\pi = 0.3$.}
     \label{fig:recovery-rate-partial}
\end{figure}

\begin{figure}
    \centering
    \includegraphics[scale = 0.8]{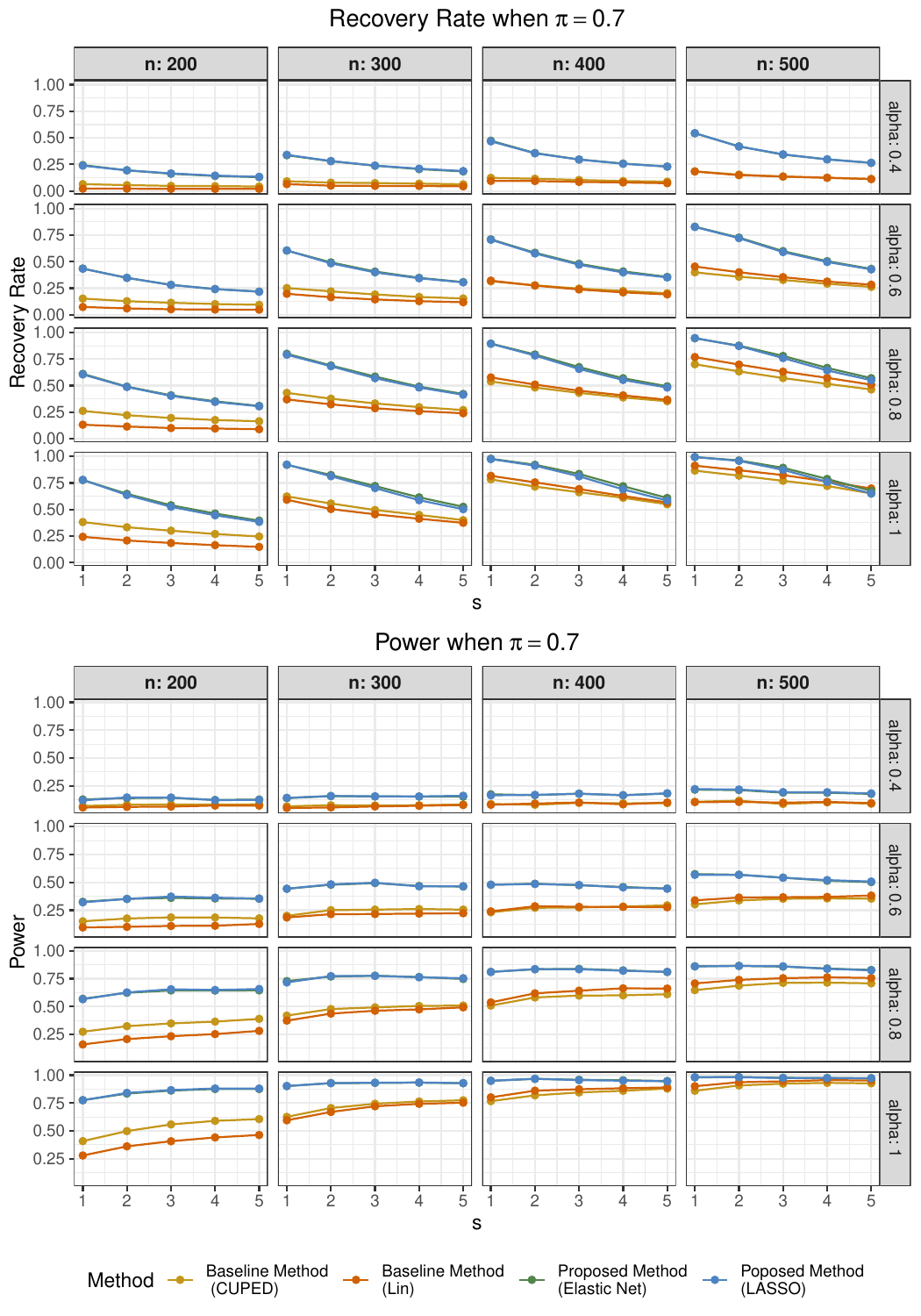}
    \caption{Recovery rates and powers averaged over 1000 replicates for the baseline and proposed approaches with pre-treatment covariates adjustment when pre-treament covariates are partially observable for $n = 200, 300, 400, 500$ and $\alpha = 0.4, 0.6, 0.8, 1$ for $\pi = 0.7$.}
     \label{fig:recovery-rate-partial-2}
\end{figure}

When pre-treament covariates are partially observable under the simulation model in Section \ref{sec:simulations}, the sparsity assumption of the inverse covariance matrix in Section \ref{sec:recovery} is violated. In the following, we add simulation results where we only observe a subset of pre-treament covariates; we only observe $40$ out of $m = 50$. The results are presented in Figure~\ref{fig:recovery-rate-partial} and \ref{fig:recovery-rate-partial-2}. We see that the proposed approach still mostly outperforms the baseline approach, especially when sample sizes are small and signals are weak. 

\end{document}